\documentclass[aip,jcp,reprint,floatfix]{revtex4-1}
\usepackage{amsmath,amssymb}
\usepackage{epsfig}
\usepackage{graphicx}
\usepackage{dcolumn}
\usepackage{bbold}
\usepackage{natbib}
\usepackage[normalem]{ulem}
\usepackage[colorlinks=true,citecolor={blue},urlcolor={blue}, linkcolor=blue]{hyperref}%
\usepackage{braket}

\newcommand{\dtp}{\partial^{\hspace{.03em}\prime}_t}

\begin{document}


\title{Time-dependent orbital-optimized coupled-cluster methods families for fermion-mixtures dynamics
}


\author{Haifeng Lang}
\email[Electronic mail:]{hflang@g.ecc.u-tokyo.ac.jp}
\affiliation{
Department of Nuclear Engineering and Management, Graduate School of Engineering, The University of Tokyo, 7-3-1 Hongo, Bunkyo-ku, Tokyo 113-8656, Japan
}
\author{Takeshi Sato}
\email[Electronic mail:]{sato@atto.t.u-tokyo.ac.jp}
\affiliation{
Department of Nuclear Engineering and Management, Graduate School of Engineering, The University of Tokyo, 7-3-1 Hongo, Bunkyo-ku, Tokyo 113-8656, Japan
}
\affiliation{
Photon Science Center, Graduate School of Engineering, The University of Tokyo, 7-3-1 Hongo, Bunkyo-ku, Tokyo 113-8656, Japan
}
\affiliation{
Research Institute for Photon Science and Laser Technology, The University of Tokyo, 7-3-1 Hongo, Bunkyo-ku, Tokyo 113-0033 Japan
}


\date{\today}

\begin{abstract}

Five time-dependent orbital optimized coupled-cluster (TD-ooCC) methods, of which four can converge to the complete active space self-consistent-field method, are presented for fermion-mixtures with arbitrary fermion kinds and numbers. Truncation schemes maintaining the intragroup orbital rotation invariance, as well as equations of motion of CC amplitudes and orbitals, are derived. Present methods are compact CC-parameterization alternatives to the time-dependent multiconfiguration self-consistent-field method for systems consisting of arbitrarily different kinds and numbers of interacting fermions. Theoretical analysis of applications of present methods to various chemical systems are reported.

\end{abstract}


\maketitle 


\section{introduction}\label{sec:introduction}

Simulating dynamics of fermions and fermion mixtures quantitatively is essential for understanding and predicting strong-field and ultrafast phenomena\cite{Protopapas1997RPP,Agostini2004RPP,Krausz2009RMP,Gallmann2013ARPC,Nisoli:2017} of molecular systems (Noticing that Hilbert spaces of vibrational systems and photons are isomorphic to fermion mixtures with one occupation in each kind\cite{mordovina2020polaritonic,christiansen2004vibrational,hansen2019time,christiansen2004vibrational,sverdrup2023time}). However, solving the time-dependent Schrödinger equation (TDSE) exactly is impractical for even moderate size systems due to the factorial growth of computational costs. Commonly used benchmark methods to describe such dynamics are the
multiconfiguration time-dependent Hartree-Fock (MCTDHF) method,\cite{Zanghellini:2003,Kato:2004,Caillat:2005,Nest:2005a,Kato:2008,Hochstuhl:2011,Haxton:2012,Haxton:2014} 
as well as its extensions for fermion-mixtures\cite{lode2020colloquium,alon2007unified}. In these methods, the wavefunction is expanded as a full configuration-interaction (CI) form inside an active space, of which single particle orbitals are time-dependent. Equations of motion (EoMs) of parameters are determined by time-dependent variational principle (TDVP). Due to the compactness of the ansatz, the numbers of optimized time-dependent orbitals are usually much less than the numbers of grid points in TDSE, and thus reducing the computational cost hugely. However, the full CI expansion makes methods still suffer from the factorial growth of the computational cost with respect to the number of fermions. Therefore, the methods can be only applied to moderate size systems. The more general time-dependent multiconfiguration self-consistent-field (TD-MCSCF) method\cite{Nguyen-Dang:2007,Ishikawa:2015,Anzaki:2017} further restricts configurations in the wavefunction expansion, which bridges the gap bewteen time-dependent Hartree-Fock and MCTDHF. One of the attractive TD-MCSCF methods, the time-dependent complete-active-space self-consistent-field (TD-CASSCF) method \cite{Sato:2013,li2021implementation}, significantly reduces the cost by assigning some occupied orbitals as cores (which must be occupied in all configurations within the expansion), resulting in a computational cost that scales factorially with respect to the number of fermions in the active orbitals rather than the total number of fermions. It is also possible to restrict configurations in the active space by the excitation levels with respect to the reference state, for instance, MCTDH[n]\cite{madsen2020systematic} and multireference MCTDH[n]\cite{madsen2020mr} for the vibrational dynamics, and time-dependent restricted active space self-consistent field (TD-RASSCF)\cite{Miyagi:2013,Miyagi:2014b,Haxton:2015} and time-dependent occupation-restricted multiple active space (TD-ORMAS)\cite{Sato:2015} for the electron dynamics, to achieve the polynomial cost scaling\cite{Haxton:2015,Sato:2016,Sawada:2016,Omiste:2017} with a drawback of not being size extensive\cite{Helgaker:2002}.

Recently, the polynomial scaling and size-extensive CC parameterization\cite{Szabo:1996,Helgaker:2002,Kummel:2003,Shavitt:2009} in the time-dependent active space gains lots of popularity. In the CC parameterization, the ket vector is described by an exponential truncated excitation operator acting on the reference state, while the bra vector is not the Hermitian conjugate of the ket vector. For this reason, the time-dependent bivariational principle\cite{Arponen:1983} and biorthogonal orbitals are also widely used in additional to the TDVP and orthogonal orbitals. Many methods have been developed and successively implemented, e.g., time-dependent optimized coupled-cluster method (TD-OCC)\cite{sato2018communication,pathak2020time1,pathak2020time2,pathak2021time} and orbital-adapted time-dependent coupled-cluster\cite{Kvaal:2012} (OATDCC) method for the electron dynamics, and time-dependent vibrational coupled cluster with time-dependent modals (TDMVCC) families\cite{madsen2020time,hojlund2024time,hojlund2024bivariational,sverdrup2023time,hojlund2022bivariational,jensen2023efficient} for the vibrational dynamics. These methods can be considered as the time-dependent extensions of stationary OCC\cite{Scuseria:1987,Sherrill:1998,Krylov:1998,Kohn:2005} and nonorthogonal orbital optimized coupled cluster (NOCC)\cite{pedersen2001gauge}.

Despite the fruitful results of CC parameterization of the active space wavefunction, it is still less explored than the CI parameterization. First, previous works mainly focus on either electron dynamics (one kind fermion) or vibrational dynamics (fermion mixtures but only one occupation in each fermion kind). A general theory of arbitrary occupations of fermion mixtures is still unclear. Second, numerically stable and convergent\cite{Kohn:2005,myhre2018demonstrating} (in the sense of converging to the CASSCF/TD-CASSCF) CC parameterizations using orthogonal orbitals, for instance, the standard CC\cite{Huber:2011,Nascimento:2016} and Brueckner coupled cluster (BCC)\cite{raghavachari1990size,hampel1992comparison,chiles1981electron,handy1989size} types of parameterization in the active space, are unknown. In this paper, we advance in these directions by presenting five time-dependent orbital-optimized coupled-cluster (TD-ooCC) methods for a system composed of arbitrarily different kinds and numbers of interacting fermions (with particle conservation).

This paper is organized as follows. In the theory section Sec.~\ref{sec:general}, we first analyze the convergence condition of time-dependent CC with time varying orthogonal orbitals to the TD-CASSCF. Further, the selections of both single CC (de)exication operators ($\hat{T}_1$ and $\hat{\Lambda}_1$, defined in Eq.~(\ref{eq:Tdefinition},\ref{eq:Lambdadefinition})), which determines the type of the method, and non-single CC (de)exication operators ($\hat{T}_0$, $\hat{T}_1$, $\cdots$, $\hat{\Lambda}_0$ $\hat{\Lambda}_2$, $\cdots$, defined in Eq.~(\ref{eq:Tdefinition},\ref{eq:Lambdadefinition})) , which can reduce the computational resources potentially, in the wavefunction ansatz of TD-ooCC methods are introduced. The equations of motion (EoMs) for CC amplitudes and orbitals are presented. Finally, we discuss their advantages and potential drawbacks. In the application section Sec.~\ref{sec:application}, we discuss three major applications of our methods to the chemical systems, electronic dynamics, vibrational dynamics, and non-adiabatic dynamics. 
Sec.~\ref{sec:sum} summarizes our theories. The Hartree atomic units are used throughout unless otherwise noted.

\section{Theory\label{sec:general}}%

\subsection{Full coupled cluster expansion in the complete active space}

Following the CASSCF method, general orthogonal time-dependent orbitals $\mu^m,\nu^m,\cdots$ of fermion kind $m$ are classified into virtual orbitals $\alpha^m,\beta^m,\cdots$, and occupied orbitals $\bar{p}^m,\bar{q}^m,\cdots$, which can be further classified into frozen-cores $i^{\prime\prime m},j^{\prime\prime m}\cdots$, dynamical-cores $i^{\prime m},j^{\prime m}$, and active orbitals $t^m,u^m,\cdots$ that contain hole orbitals $i^m,j^m,\cdots$, and particle orbitals $a^m,b^m,\cdots$. Here, the hole orbitals and particle orbitals refer to active orbitals that are occupied and unoccupied, respectively, in the reference state. EoMs of orbitals can be formally expressed as
\begin{equation}
i|\dot{\psi}_{\nu^m}\rangle = 
  i|\psi_{\mu^m}\rangle X^{\mu^m}_{\nu^m} \,,\label{eq:eom_orb} 
\end{equation}
where $X^{\mu^m}_{\nu^m} = \braket{\psi_{\mu^m}|\dot{\psi}_{\nu^m}}$ is anti-Hermitian to enforce the orthonormality of orbitals $\langle\psi_{\mu^m}(t)|\psi_{\nu^m}(t)\rangle=\delta^{\mu^m}_{\nu^m}$ at any time, and thus called the constraints of orbital rotations. There are two types of constraints of orbitals: the first type includes pre-determined constraints that are intra-group rotation constraints and frozen-core related orbital constraints\cite{Constraints}; the second type includes others that are variational parameters which should be determined by the time-dependent variational principle. In this article, we will use the Einstein notation except for fermion kind index $m,n,\cdots$. For the convenience, we use the notations $p^m,q^m,\cdots$ to label time-dependent occupied orbitals except for the frozen-cores due to the fact that frozen-cores are pre-determineded by the gauge choice\cite{Sato:2016}. Additionally, $\hat{c}_{\mu^m}^\dagger$ and $\hat{c}_{\mu^m}$ are defined as the creation and annihilation operators for $\psi_{\mu^m}$ (as well as other orbital labels).

The most general ansatz of the CC wave function with time-dependent orbital can be expressed as
\begin{equation}
    \ket{\Psi_{\rm R}} = e^{\hat{T}}\ket{\Phi}\,,\quad \bra{\Psi_{\rm L}} = \bra{\Phi}\hat{\Lambda}e^{-\hat{T}}\,,
\end{equation}
where $\ket{\Phi}$ and $\bra{\Phi}$ are the reference states, i.e., tensor product of Slater determinants of all fermion kinds $\ket{\Phi} = \prod_m\prod_{i^{\prime\prime m}i^{\prime m}i^m} \hat{c}^\dagger_{i^{\prime\prime m}}\hat{c}^\dagger_{i^{\prime m}}\hat{c}^\dagger_{i^{ m}}\ket{0}$, and
\begin{equation}\label{eq:Tdefinition}
    \begin{aligned}
        \hat{T} &= \hat{T}_0 + \hat{T}_1 + \hat{T}_2+ \hat{T}_3+ \cdots \\ &= \tau_0\hat{E}_0 + \sum_{m}\tau_{i^m}^{a^m} \hat{E}_{i^m}^{a^m} +
\sum_{m_1m_2}\tau_{i^{m_1}j^{m_2}}^{a^{m_1}b^{m_2}} \hat{E}_{i^{m_1}j^{m_2}}^{a^{m_1}b^{m_2}}  \\
&+\sum_{m_1m_2m_3}\tau_{i^{m_1}j^{m_2}k^{m_3}}^{a^{m_1}b^{m_2}c^{m_3}}\hat{E}_{i^{m_1}j^{m_2}k^{m_3}}^{a^{m_1}b^{m_2}c^{m_3}} + \cdots \\ &= \tau^{\mathring{A}}_{\mathring{I}} \hat{E}^{\mathring{A}}_{\mathring{I}} =
 \sum_{m}\tau_{i^m}^{a^m} \hat{E}_{i^m}^{a^m} + \tau^{{A}}_{{I}} \hat{E}^{{A}}_{{I}} \,,
    \end{aligned}
\end{equation}
\begin{equation}\label{eq:Lambdadefinition}
    \begin{aligned}
        \hat{\Lambda} & = \hat{\Lambda}_0 + \hat{\Lambda}_1 + \hat{\Lambda}_2+ \hat{\Lambda}_3+ \cdots \\ &= \lambda_0\hat{E}_0 + \sum_{m}\lambda^{i^m}_{a^m} \hat{E}^{i^m}_{a^m} +
\sum_{m_1m_2}\lambda^{i^{m_1}j^{m_2}}_{a^{m_1}b^{m_2}} \hat{E}^{i^{m_1}j^{m_2}}_{a^{m_1}b^{m_2}}  \\
&+\sum_{m_1m_2m_3}\lambda^{i^{m_1}j^{m_2}k^{m_3}}_{a^{m_1}b^{m_2}c^{m_3}}\hat{E}^{i^{m_1}j^{m_2}k^{m_3}}_{a^{m_1}b^{m_2}c^{m_3}} + \cdots \\ &= \lambda_{\mathring{A}}^{\mathring{I}} \hat{E}_{\mathring{A}}^{\mathring{I}} 
=\sum_{m}\lambda^{i^m}_{a^m} \hat{E}^{i^m}_{a^m} + \lambda_{{A}}^{{I}} \hat{E}_{{A}}^{{I}} \,,
    \end{aligned}
\end{equation}
Here, $\mathring{I}=ijk\cdots$ and $\mathring{A}=abc\cdots$ (can be empty) denote hole and particle orbital strings to make the notation brevity. The general-rank excitation (deexcitation) operators and amplitudes are $\hat{E}^{\mathring{A}}_{\mathring{I}}$ ($\hat{E}_{\mathring{A}}^{\mathring{I}}$) and $\tau^{\mathring{A}}_{\mathring{I}}$ ($\lambda_{\mathring{A}}^{\mathring{I}}$), where ${\hat{E}}^{\mu_1^{m_1}\mu_2^{m_2}\mu_3^{m_3}\cdots}_{\nu_1^{m_1}\nu_2^{m_2}\nu_3^{m_3}\cdots}=\hat{c}^\dagger_{\mu_1^{m_1}}\hat{c}^\dagger_{\mu_2^{m_2}}\hat{c}^\dagger_{\mu_3^{m_3}}\cdots{\hat{c}}_{\nu_3^{m_3}}{\hat{c}}_{\nu_2^{m_2}}{\hat{c}}_{\nu_1^{m_1}}$, and $\hat{E}_0 = \hat{I}$ for the zero-rank excitation (empty strings of $\mathring{I}$ and $\mathring{A}$). We will also use the notation $I$ and $A$ to label non-single hole and particle orbital strings (the orbital number of the strings is not one) in this article.

Before working on time-dependent variational principles, we stress that single amplitudes terms $\hat{T}_1$ and $\hat{\Lambda}_1$ are included in the ansatz due to the convenience of the theoretical analysis although they can cause the numerical instability in practical simulations and are omitted in previous work\cite{sato2018communication,Kohn:2005}. Additionally, zero-rank amplitudes $\tau_0$ and $\lambda_0$ are also explicitly included in the ansatz to recover the CASSCF wavefucntion in the full CC expansion limit. The normalization condition makes $\lambda_0\equiv 1$ conserved, and $\tau_0$ has no impacts on observables but essential for the autocorrelations.

Following the previous work\cite{Pedersen:1999,Kvaal:2012,Helgaker:2002,sato2018communication}, we begin with the
coupled-cluster Lagrangian $L$ and real action functional $S$,
\begin{align}\label{eq:lagrangian}
L &= \langle\Psi_{\rm L}|(\hat{H}-i{\partial_t})|\Psi_{\rm R}\rangle\,,\\
S &= \label{eq:action}
\Re\int_{t_0}^{t_1} Ldt = \frac{1}{2} \int_{t_0}^{t_1}\left( L + L^*\right) dt\,.
\end{align}
Here, the Hamiltonian $\hat{H}$ can be an arbitrary Hamiltonian with particle conservations. EoMs of parameters are given by the stationary condition of the action with respect to the variation of all parameters of left- and right-state {\color{black}wavefunctions} (with necessary constraints),

\begin{widetext}
{\color{black}
\begin{eqnarray} \label{eq:vars}
{\color{black}2}\delta S 
&=&
\delta\tau^{\mathring{A}}_{\mathring{I}}\left\{
\langle\Psi_{\rm L}|
[\hat{H}-i\hat{X},\hat{E}^{\mathring{A}}_{\mathring{I}}]|\Psi_{\rm R}\rangle
+i\dot{\lambda}_{\mathring{A}}^{\mathring{I}}\right\} 
+\delta\lambda_{\mathring{A}}^{\mathring{I}}\left\{
\langle\Phi^{\mathring{A}}_{\mathring{I}}|e^{-\hat{T}}
(\hat{H}-i\hat{X})|\Psi_{\rm R}\rangle
-i\dot{\tau}^{\mathring{A}}_{\mathring{I}}\right\} + c.c \nonumber \\
&+& \sum_m
\Delta^{\mu^m}_{\nu^m} \left\{
\langle\Psi_{\rm L}|[\hat{H}-i(\dtp\hat{T})-i\hat{X},\hat{E}^{\mu^m}_{\nu^m}]|\Psi_{\rm R}\rangle
+i\langle\Phi|(\dtp\hat{\Lambda})e^{-\hat{T}}\hat{E}^{\mu^m}_{\nu^m}|\Psi_{\rm R}\rangle
\right. \nonumber \\
&& \label{eq:vars_orb}
\hspace{1.3em} \left.-
\langle\Psi_{\rm L}|[\hat{H}-i(\dtp\hat{T})-i\hat{X},\hat{E}_{\mu^m}^{\nu^m}]|\Psi_{\rm R}\rangle^*
+i\langle\Phi|(\dtp\hat{\Lambda})e^{-\hat{T}}\hat{E}_{\mu^m}^{\nu^m}|\Psi_{\rm R}\rangle^*
\right\}\,.
\end{eqnarray}
}
\end{widetext}
where 
$\dtp\hat{T}=\dot{\tau}^{\mathring{A}}_{\mathring{I}}\hat{E}^{\mathring{A}}_{\mathring{I}}$,
$\dtp\hat{\Lambda}=\dot{\lambda}_{\mathring{A}}^{\mathring{I}}\hat{E}_{\mathring{A}}^{\mathring{I}}$,
$\hat{X}=\sum_mX^{\mu^m}_{\nu^m}\hat{E}^{\mu^m}_{\nu^m}$,
$\hat{\Delta}=\sum_m\Delta^{\mu^m}_{\nu^m}\hat{E}^{\mu^m}_{\nu^m}$, and
$\Delta^{\mu^m}_{\nu^m} =  \braket{\psi_{\mu^m}|\delta{\psi}_{\nu^m}}$.
Similar to the matrices  $X^{\mu^m}_{\nu^m}$, $\Delta^{\mu^m}_{\nu^m} $
are also anti-Hermitian to enforce the orthonormality of orbitals. Meanwhile, $\Delta^{\mu^m}_{i^{\prime \prime m}}\equiv 0$ due to the fact that EoMs of frozen-cores are not determined by variational principle. Additionally, we only consider the  ansatz that satisfies the intragroup rotation invariance including dynamical-cores-dynamical-cores, hole-hole, particle-particle, and virtual-virtual rotation invariance, i.e., the wavefunction is invariant up to a phase factor with these orbital rotations and the corresponding inverse rotations of CC amplitudes. These invariances give the redundancy of $X^{i^{\prime  m}}_{j^{\prime  m}}$, $X_{i^m}^{j^m}$, $X_{a^m}^{b^m}$, and $X_{\alpha^m}^{\beta^m}$. For simplicity, we always set all of them as zero. The details of the wavefunction ansatz that maintain the hole-hole, particle-particle, and virtual-virtual rotation invariance will be discussed in the Sec.~\ref{sec.truncation} and ~\ref{Sec.Redundancy}.

Variations with respect to CC amplitudes, $\delta S/\delta \lambda_{\mathring{A}}^{\mathring{I}} =0$ and $\delta S/\delta \tau^{\mathring{A}}_{\mathring{I}}=0$ give 
\begin{align}
    i\dot{\tau}^A_I &= \label{eq:eom_t0}\langle\Phi^{{A}}_{{I}}| e^{-\hat{T}} (\hat{H}-i\hat{X})|\Psi_{\rm R}\rangle\,, \\
    i\dot{\tau}_{i^m}^{a^m} &=  \bra{\Phi_{i^m}^{a^m}}e^{-\hat{T}}(\hat{H} - i\hat{X})\ket{\Psi_{\rm R}}\,,\label{eq:eom_t1} \\
     -i\dot{\lambda}^I_A &=
\langle\Psi_{\rm L}| e^{-\hat{T}} [\hat{H}-i\hat{X},\hat{E}^A_I] |\Psi_{\rm R}\rangle\,,  \label{eq:eom_l0}\\
-i\dot{\lambda}^{i^m}_{a^m} &= \bra{\Psi_{\rm L}}[\hat{H} - i\hat{X} ,\hat{E}_{i^m}^{a^m}]\ket{\Psi_{\rm R}}\label{eq:eom_l1}\,. 
\end{align}
Here, we separate EoMs of CC amplitudes into non-single amplitudes part and single amplitudes part. The variation with respect to $\Delta_{a^m}^{i^m}$ gives
\begin{equation}
\begin{aligned}
    &\bra{\Psi_{\rm L}}[\hat{H} - i\hat{X} - i(\partial_t^\prime \hat{T}),\hat{E}^{i^m}_{a^m}]\ket{\Psi_{\rm R}} \\
    + & i\bra{\Phi}(\partial_t^\prime \hat{\Lambda})e^{-\hat{T}}\hat{E}^{i^m}_{a^m}|\Psi_{\rm R}\rangle \\
    - &\bra{\Psi_{\rm L}}[\hat{H} - i\hat{X} - i(\partial_t^\prime \hat{T}),\hat{E}_{i^m}^{a^m}]\ket{\Psi_{\rm R}}^* \\
    + & i\bra{\Phi}(\partial_t^\prime \hat{\Lambda})e^{-\hat{T}}\hat{E}_{i^m}^{a^m}|\Psi_{\rm R}\rangle^* = 0\,, \label{eq:eom_X1}
\end{aligned}
\end{equation}
Substituting Eq.~(\ref{eq:eom_t0},\ref{eq:eom_l0}) into Eq.~(\ref{eq:eom_X1}) yields,
\begin{equation}
    \begin{aligned}
&i(\delta_{b^m}^{a^m}D_{i^m}^{j^m}-D_{b^m}^{a^m}\delta_{i^m}^{j^m})(\dot{\tau}_{j^m}^{b^m} +  X_{j^m}^{b^m}) +i\frac{1}{2}(\dot{\lambda}^{i^m}_{a^m})^* = \\
&B_{i^m}^{a^m} -i\frac{1}{2}\sum_n\left\{\langle\Phi_{j^n}^{b^n}| e^{-\hat{T}} \hat{E}^{i^m}_{a^m}|\Psi_{\rm R}\rangle\dot{\lambda}^{j^n}_{b^n} + A^{a^mj^n}_{i^mb^n}  (X^{b^n}_{j^n})^* \right\} \,, \label{eq:eom_Xia}
    \end{aligned}
\end{equation}
where
\begin{align}
A^{a^mj^n}_{i^mb^n} &= \label{eq:amat}
\bra{\Psi_{\rm L}} [\hat{E}^A_I,\hat{E}^{j^n}_{b^n}] \ket{\Psi_{\rm R}}
\langle\Phi^A_I|e^{-\hat{T}}\hat{E}^{i^m}_{a^m}\ket{\Psi_{\rm R}}
\nonumber \\ &-
\bra{\Psi_{\rm L}} [\hat{E}^A_I,\hat{E}^{i^m}_{a^m}]\ket{\Psi_{\rm R}}
\langle\Phi_I^A| e^{-\hat{T}} \hat{E}^{j^n}_{b^n}\ket{\Psi_{\rm R}}, \\
B_{i^m}^{a^m} &= 
-\frac{1}{2}\braket{\Psi_{\rm L}|[\hat{H},\hat{E}^{i^m}_{a^m}]|\Psi_{\rm R}} - \frac{1}{2}\braket{\Psi_{\rm R}|[\hat{H},\hat{E}^{i^m}_{a^m}]|\Psi_{\rm L}}  \nonumber\\
&-
\frac{1}{2}\bra{\Psi_{\rm L}}  [\hat{E}^A_I,\hat{H}] \ket{\Psi_{\rm R}}
\langle\Phi^A_I|e^{-\hat{T}}\hat{E}^{i^m}_{a^m}\ket{\Psi_{\rm R}}
\nonumber \\ &
+
\frac{1}{2}\bra{\Psi_{\rm L}} [\hat{E}^A_I,\hat{E}^{i^m}_{a^m}]\ket{\Psi_{\rm R}}
\langle\Phi_I^A| e^{-\hat{T}} \hat{H}\ket{\Psi_{\rm R}}\,. \label{eq:bvec}
\end{align}
Solving coupled equations Eq.~(\ref{eq:eom_t1},\ref{eq:eom_l1},\ref{eq:eom_Xia}), EoMs of single amplitudes and particle-hole orbital rotations $\dot{\tau}_{i^m}^{a^m}$, $\dot{\lambda}^{i^m}_{a^m}$, and $ X_{i^m}^{a^m}$ can be obtained. Substituting $ X_{i^m}^{a^m}$ into Eq.~(\ref{eq:eom_t0},\ref{eq:eom_l0}), EoMs of non-single amplitudes can be solved.

In the full CC expansion limit, the particle-hole orbital rotations $X_{i^m}^{a^m}$ are redundant, and can be freely selected. For an explicit proof, see the Appendix.~\ref{Sec.Redundancy}. The simplest and the most common choice is to set them as zero. However, coupling equations Eq.~(\ref{eq:eom_t1},\ref{eq:eom_l1},\ref{eq:eom_Xia}) suggest that the redundancy of $X_{i^m}^{a^m}$ can be transferred to $\lambda^{i^m}_{a^m}$ or $\tau_{i^m}^{a^m}$. For instance, one can set $\lambda_{a^m}^{i^m}\equiv 0$ as redundant parameters, and this type of wavefunction parameterization is known as the BCC\cite{raghavachari1990size,hampel1992comparison,chiles1981electron,handy1989size}. In fact, one can arbitrarily select one of $X_{i^m}^{a^m}$, $\tau_{i^m}^{a^m}$ and ${\lambda}^{i^m}_{a^m}$ as redundant parameter for each kind of fermion, and set them as zero. In the Appendix.~\ref{Sec.Redundancy}, more rigorous arguments of transferring redundancy and preparation of wavefunctions are presented.

\subsection{Wavefunction Ansatz of Time dependent orbital-optimized coupled-cluster families}
\subsubsection{Selection of single amplitudes}
In principle, the redundancy caused by including all $X_{i^m}^{a^m}$, $\tau_{i^m}^{a^m}$, and $\lambda^{i^m}_{a^m}$ only occurs at the full CC expansion limit. Nonetheless, our experience shows that including them all is numerically instable even when only double (rank two) excitations are considered, which suggests the compactness of CC wavefunction parameterization. One has to exclude (at least) one of $X_{i^m}^{a^m}$, $\tau_{i^m}^{a^m}$ and $\lambda^{i^m}_{a^m}$ in practical simulations.

The first method we will consider is TD-OCC (terminology adapted from the reference\cite{sato2018communication}), of which the ansatz only includes $X_{i^m}^{a^m}$. This method fails to converge to the CASSCF limit\cite{Kohn:2005,myhre2018demonstrating,hojlund2024time}, but shows an acceptable accuracy in ultrafast electron dynamics\cite{sato2018communication} and vibrational systems\cite{hojlund2024time} according to previous reports. Specifically, TD-OCC with double and triple excitations (TD-OCCDT) is in nearly perfect agreement with the CASSCF results for benchmarks of ultrafast electron dynamics\cite{sato2018communication}. We will also present three methods with the absence of one of $X_{i^m}^{a^m}$, $\tau_{i^m}^{a^m}$, and $\lambda^{i^m}_{a^m}$ in the ansatz, thus all three methods can converge to the CASSCF limit. The details of four methods are summarized as:

(a). TD-OCC: only $X_{i^m}^{a^m} $ are included in the ansatz, or equivalently $\tau_{i^m}^{a^m}=\lambda^{i^m}_{a^m} = \delta\tau_{i^m}^{a^m} = \delta\lambda^{i^m}_{a^m}  \equiv 0$.

(b). TD-OCCT1: $\tau_{i^m}^{a^m}$ and $X_{i^m}^{a^m}$ are included in the ansatz, or equivalently $\lambda^{i^m}_{a^m}= \delta\lambda^{i^m}_{a^m}\equiv 0$.

(c). TD-BCC: $X_{i^m}^{a^m}$ and $\lambda^{i^m}_{a^m}$ are included in the ansatz, or equivalently $\tau_{i^m}^{a^m}=\delta\tau_{i^m}^{a^m} \equiv 0$.

(d). TD-OCCX0: $\tau_{i^m}^{a^m}$ and $\lambda^{i^m}_{a^m}$ are included in the ansatz, or equivalently $X_{i^m}^{a^m}=\Delta_{i^m}^{a^m} \equiv 0$.

Here, the latter definition of each method can be regarded as the constraints in the sense that some parameters are pre-determined (to be zero). We stress that the absence of particle-hole orbital rotations $X_{i^m}^{a^m}$ of TD-OCCX0 is only justified for the dynamics without external fields. In the presence of external fields $X_{i^m}^{a^m}$ has to be assigned different values to ensure the gauge invariance,  akin to the treatment of frozen-cores\cite{Sato:2016}. Additionally, we will consider the most general hybrid form of wavefunction parameterization, i.e., the parameter selections (a), (b), (c), or (d) is separately, and flexibly adopted for each fermion kinds, which is named as TD-OCCH and will be discussed with all the details in the Appendix.\ref{Sec.TD-OCCH}.

\subsubsection{Selection of non-single excitations}\label{sec.truncation}
Inspired by TD-MCSCF, one can restrict the non-single excitations in TD-ooCC to make the ansatz more compact. To ensure the intragroup rotation invariance of particle and hole orbitals, one can select non-single excitations that satisfy the property
\begin{equation}\label{eq:criteria}
    \begin{aligned}
         &\hat{E}^{\mu^m}_{\nu^m}\ket{\Psi_{\rm R}} = e^{\hat{T}}\hat{\Pi}e^{-\hat{T}}\hat{E}^{\mu^m}_{\nu^m}\ket{\Psi_{\rm R}} \,,\\ 
 &\bra{\Psi_{\rm L}}\hat{E}^{\mu^m}_{\nu^m} = \bra{\Psi_{\rm L}}\hat{E}^{\mu^m}_{\nu^m} e^{\hat{T}}\hat{\Pi}e^{-\hat{T}} \,,
    \end{aligned}
\end{equation}
where $\hat{\Pi} = \ket{\Phi_{\mathring{I}}^{\mathring{A}}}\bra{\Phi_{\mathring{I}}^{\mathring{A}}}$. Here $\mu^m$ and $\nu^m$ should be particle or hole orbital index simultaneously. For the details of the proof, see the Appendix.~\ref{Sec.Redundancy}. One of the simplest selection scheme is the $[k]$-truncation scheme, i.e., non-single excitation is included in the ansatz if and only if it is not larger than the certain truncation order $k$, which has been successfully implemented in TD-OCC.

It is also possible to assign different truncation orders for different kinds to adapt their contribution to the dynamics. Here, we propose two possible schemes. The first, the weighted excitation truncation scheme, is the modified version of the truncation scheme reported in Ref.~[\onlinecite{madsen2020general}]. Define the nonnegative excitation weight and the excitation number of each kind as $w^m$ and $n_{\rm ext}^m$, respectively. The generalized excitation level of a given excitation amplitude is 
\begin{equation}
    w = \sum_{m} n_{\rm ext}^mw^m\,,
\end{equation}
and all non-single excitation amplitude that satisfy $w \le k$ should be included in the ansatz, where $k$ is the maximum generalized excitation level. Here, different spin fermion configurations can be regarded as different ``fermion kinds" for both unrestricted and restricted spin orbitals if the each spin component is conserved. This truncation scheme can reduce to the scheme in Ref.~[\onlinecite{madsen2020general}] in the vibrational system, in which $n_{\rm ext}^m$ can only be 1 or 0. One can set the weights of important kinds as relatively small numbers, even zero. When all $w^m = 1$, the weighted excitation truncation scheme reduces to the $[k]$-truncation scheme.

The second one, group excitation truncation scheme, is the following: (1) all kinds of fermions can be divided into several groups A,B,C,$\cdots$, and their truncation order of excitations can be assigned as $n^A_A$, $n^B_B$, $n^C_C$,$\cdots$. Here, $n=1$ if only zero excitations (and single excitations when included) are included in the ansatz. All possible non-single excitations beyond the truncation order must be excluded while all others must be included in the ansatz. The different spin components of a fermion can also be regarded as different kinds if they are conserved; (2) for each type of excitations across groups, truncation order of the total excitation and all single groups should be assigned, for instance, $(n_{AB},n_{AB}^A,n_{AB}^B)$ should be assigned for the excitation acrossing groups $A$ and $B$, and $(n_{ABC}, n_{ABC}^A, n_{ABC}^B, n_{ABC}^C)$ should be assigned for the excitation acrossing groups $A$, $B$ and $C$. Iff both the total excitation number and the excitation number of each group are not larger than the corresponding truncation parameters, it is included in the ansatz; (3) $n$ must satisfy the inequalities: for all $X$, $n^X_{Y} \leq n^X_{Z}$ if the group numbers of $Y$ is larger than $Z$. We will list two explicit examples on the electron dynamics. We set the $\alpha$ spin and $\beta$ spin are two different groups. The first example is $n_\alpha^\alpha = n_\beta^\beta = n_{\alpha\beta}^\alpha = n_{\alpha\beta}^\beta = 2$ and $n_{\alpha\beta} = 3$, which means all double excitations of $\alpha$ and $\beta$ spins are included, and all $2\alpha 1\beta$ and $2\beta 1\alpha$ type triple excitations are included. The second is $n_\alpha^\alpha = n_\beta^\beta = n_{\alpha\beta}^\alpha = n_{\alpha\beta}^\beta = 1$, and $n_{\alpha\beta} = 2$, which means only $1\alpha 1\beta$ excitations are included. For a slightly more complicated example, see Fig.~(1).

 \begin{figure}[!b]
\centering
\includegraphics[width=0.9\linewidth,clip]{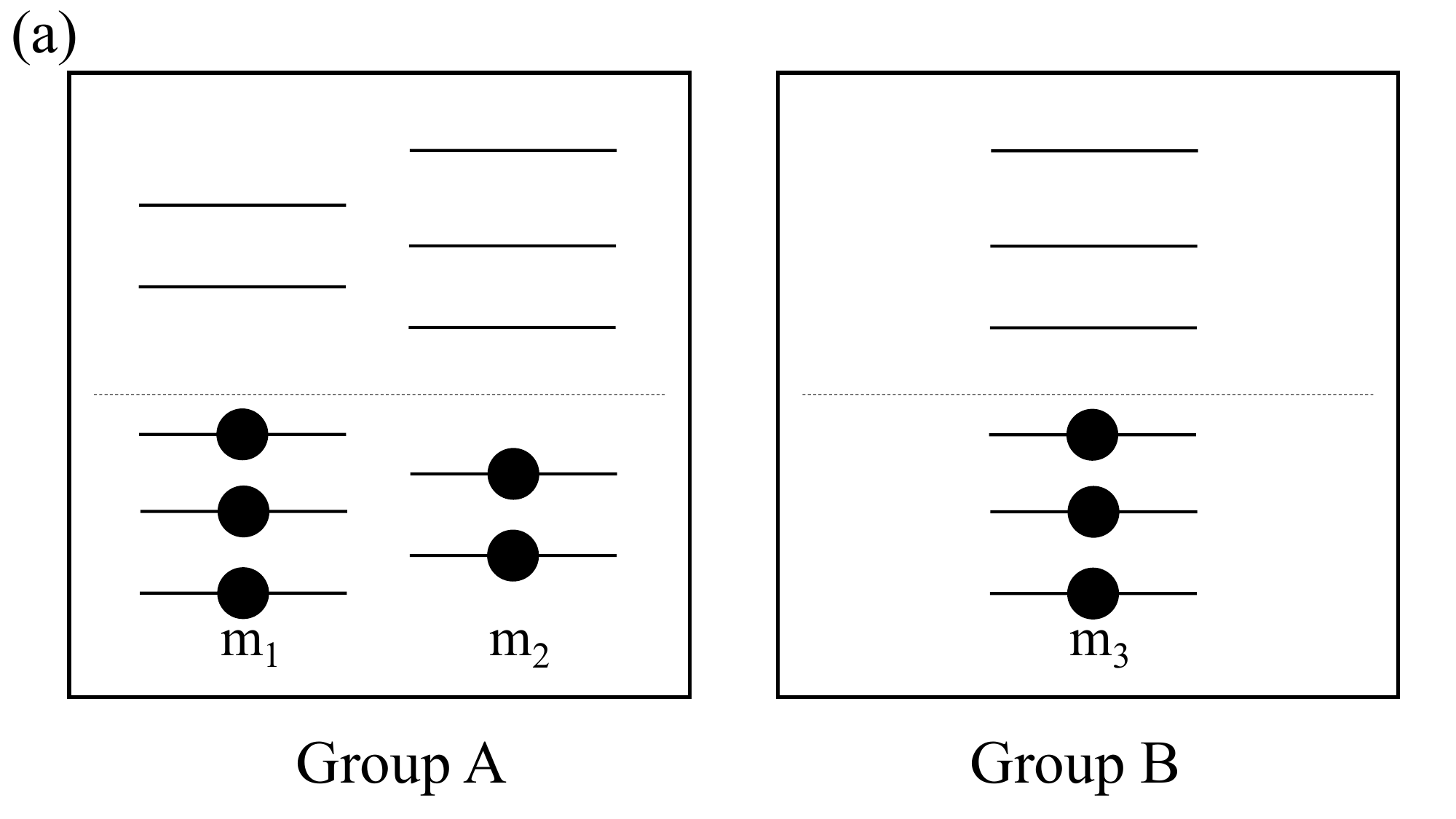}
\includegraphics[width=0.9\linewidth,clip]{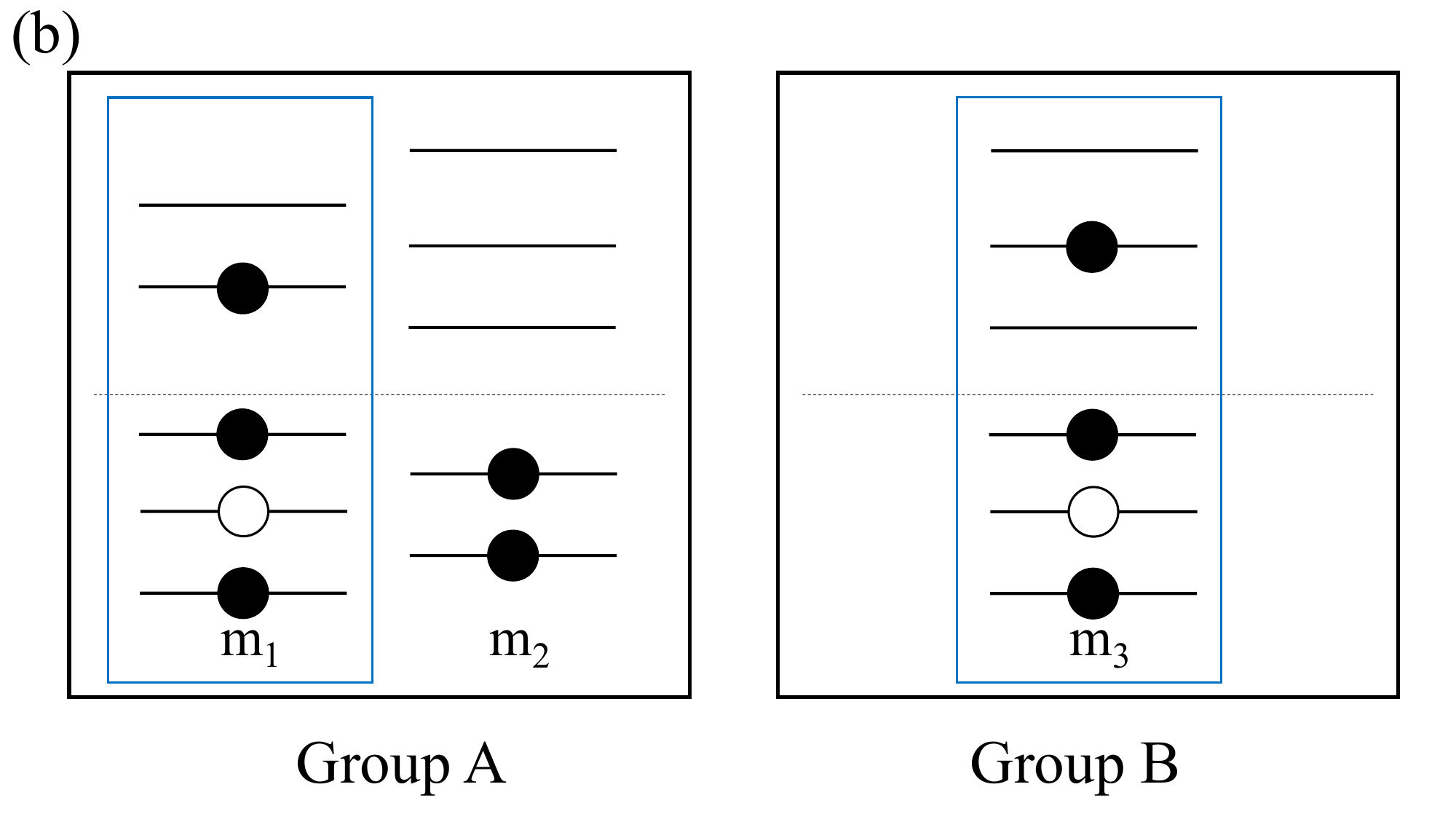}
\includegraphics[width=0.9\linewidth,clip]{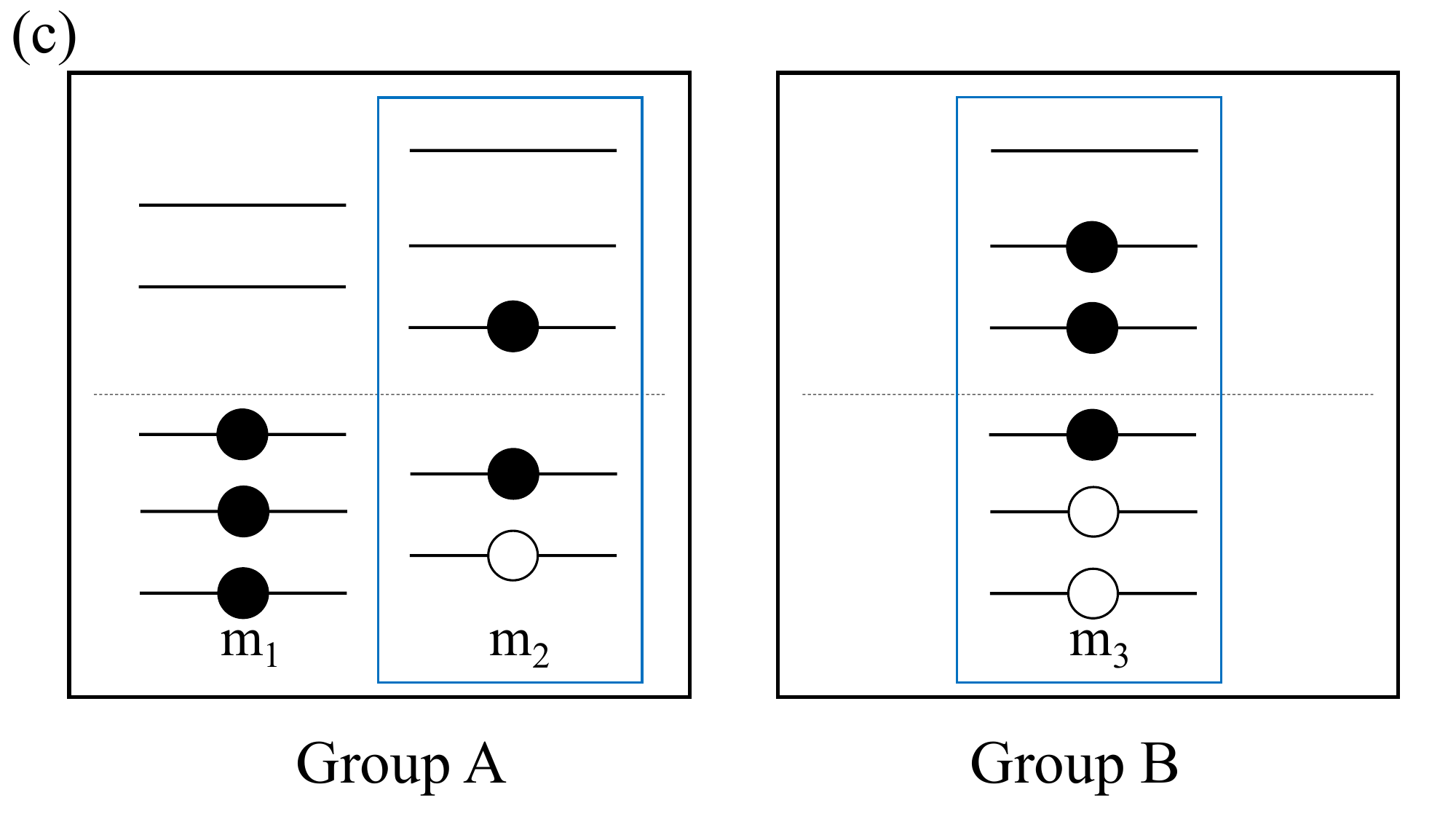}
\includegraphics[width=0.9\linewidth,clip]{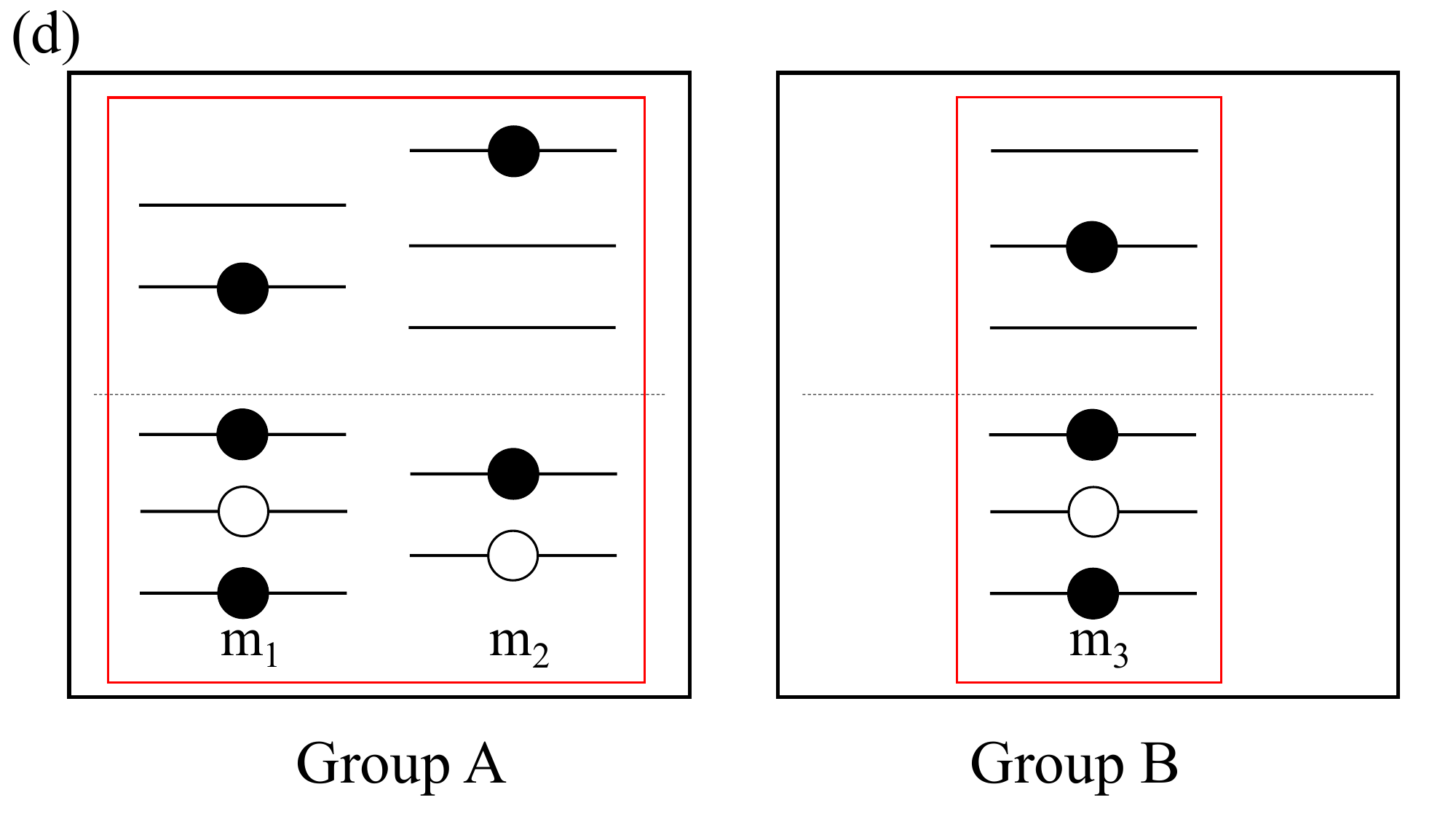}

\caption{\label{fig:TruncationExcitaion}
Illustration of the Truncation of non-single CC amplitudes (cores and virtual orbitals are neglected). In the example of this figure, three kinds fermions ($m_1$, $m_2$, and $m_3$) are classified into two groups: Group A containing $m_1$ and $m_2$, and Group B containing $m_3$. We set $n_A^A=n_B^B=n_{AB}^B=2$, $n_{AB}=3$, and $n_{AB}^A=1$. Fig.~(1a) is the reference configuration, where dashed line is the fermion surface. Figs.~(1b,c) show two of allowed CC excitations across the group, where blue rectangles circle the excitation kinds. Fig.~(1d) shows the forbidden excitations, where red rectangles circle the excitation kinds. In the Fig.~(1d), although the total excitation and excitation of group B satisfy the conditions, $3 = n_{AB}$ and $1 < n_{AB}^B$, the excitation is still forbidden due to the condition violation of the group A excitation $2>n_{AB}^A$.
}%
\end{figure}
\subsection{Equation of motions of Time dependent optimized coupled cluster families}\label{Sec.EoMs}

%
EoMs of single amplitudes and particle-hole orbital rotations of five methods can be obtained by setting corresponding constraints on Eq.~(\ref{eq:eom_t1},\ref{eq:eom_l1},\ref{eq:eom_Xia}), which will be presented in the corresponding subsubsections and the Appendix. EoMs of other variables are the same for all five methods [Eq.~(\ref{eq:eom_t0},\ref{eq:eom_l0}) ]. EoMs of intergroup rotations $X^{\alpha^m}_{p^m}$ and $X^{t^m}_{i^{\prime m}}$ are obtained from the variation with respect to $\Delta_{\alpha^m}^{p^m}$ and $\Delta_{t^m}^{i^{\prime m}}$,
\begin{equation}\label{eq:EoMstaulambda1}
    \begin{aligned}
        &\bra{\Psi_{\rm L}}[\hat{H} - i\hat{X},\hat{E}_{\alpha^m}^{p^m}]\ket{\Psi_{\rm R}} + \bra{\Psi_{\rm R}}[\hat{H} - i\hat{X},\hat{E}_{\alpha^m}^{p^m}]\ket{\Psi_{\rm L}} =0\,,\\
        &\bra{\Psi_{\rm L}}[\hat{H} - i\hat{X},\hat{E}_{t^m}^{i^{\prime m}}]\ket{\Psi_{\rm R}} + \bra{\Psi_{\rm R}}[\hat{H} - i\hat{X},\hat{E}_{t^m}^{i^{\prime m}}]\ket{\Psi_{\rm L}} =0\,,
    \end{aligned}
\end{equation}
which can be further simplified as
\begin{align}
2iX^{u^m}_{i^{\prime m}}&( 2\delta_{u^m}^{t^{ m}} - D_{u^m}^{t^{ m}} ) = \nonumber \\
&\braket{\Psi_{\rm L}|\hat{E}_{t^m}^{i^{\prime m}}\hat{H}|\Psi_{\rm R}} + \braket{\Psi_{\rm R}|\hat{E}_{t^m}^{i^{\prime m}}\hat{H}|\Psi_{\rm L}} \,, \\
2iX^{\alpha^m}_{q^{ m}} &D^{q^m}_{p^m} = \braket{\Psi_{\rm L}|\hat{E}_{\alpha^m}^{p^{ m}}\hat{H}|\Psi_{\rm R}} + \braket{\Psi_{\rm R}|\hat{E}_{\alpha^m}^{p^{ m}}\hat{H}|\Psi_{\rm L}} \,, 
\end{align}
with
\begin{align}
 \rho_{q^m}^{p^m} &= \label{eq:1rdm}
\bra{\Psi_{\rm L}}\hat{E}^{q^m}_{p^m}\ket{\Psi_{\rm R}}\,,\hspace{.5em}D_{q^m}^{p^m}=\frac{1}{2}(\rho_{q^m}^{p^m}+(\rho^{q^m}_{p^m})^*)
\end{align}
In practical simulations, virtual orbitals are never propagated. Instead, one can convert\cite{beck2000multiconfiguration,Anzaki:2017} the propagation of occupied orbitals arisen from the virtual-occupied orbital rotations to 
\begin{equation}
\begin{aligned}
    &i\braket{\psi_{\bar{\mu}^m}|\psi_{\alpha^m}}X^{\alpha^m}_{p^{ m}} = (D^{-1})_{q^m}^{p^m} \\
    \times&\Big\{  \braket{\Psi_{\rm L}|\hat{E}_{\bar{\mu}^m}^{q^{ m}}\hat{H}|\Psi_{\rm R}} 
    - \braket{\Psi_{\rm L}|\hat{E}_{\bar{o}^m}^{q^{ m}}\hat{H}|\Psi_{\rm R}}\braket{\psi_{\bar{\mu}^m}|\psi_{\bar{o}^m}} \\
    &+ \braket{\Psi_{\rm R}|\hat{E}_{\bar{\mu}^m}^{q^{ m}}\hat{H}|\Psi_{\rm L}} 
    - \braket{\Psi_{\rm R}|\hat{E}_{\bar{o}^m}^{q^{ m}}\hat{H}|\Psi_{\rm L}}\braket{\psi_{\bar{\mu}^m}|\psi_{\bar{o}^m}}  \Big\} \,,
\end{aligned}    
\end{equation}
where $\bar{\mu}^m$ is the primary time-independent basis to expand orbitals that can be discrete variables representation\cite{beck2000multiconfiguration}, grid points\cite{li2021implementation}, etc.

\subsubsection{TD-OCC}
Setting $\tau_{i^m}^{a^m}={\lambda}^{i^m}_{a^m}\equiv 0$ in Eq.~(\ref{eq:eom_Xia}),
\begin{equation}\label{eq:eom_Xia_TDOCC}
i(\delta_{b^m}^{a^m}D_{i^m}^{j^m}-D_{b^m}^{a^m}\delta_{i^m}^{j^m})X_{j^m}^{b^m}  + i\sum_n\frac{1}{2}A^{a^mj^n}_{i^mb^n}  (X^{b^n}_{j^n})^* = 
B_{i^m}^{a^m}  \,. 
\end{equation}
In the case of one occupation of each kind of fermion and a single kind of fermion, the method reduces to oTDMVCC\cite{hojlund2024time} for vibrational systems and TD-OCC\cite{sato2018communication} for electrons, respectively.

\subsubsection{TD-OCCT1}

Setting ${\lambda}^{i^m}_{a^m}\equiv 0$ in Eq.~(\ref{eq:eom_l1},\ref{eq:eom_Xia}),
\begin{equation}\label{eq:eom_Xia_TDOCCT1}
i\left( \rho_{j^m}^{i^m}\delta_{b^m}^{a^m} - \rho_{a^m}^{b^m}\delta_{j^m}^{i^m} \right)X^{j^m}_{b^m} =
\bra{\Psi_{\rm L}} [\hat{H},\hat{E}_{i^m}^{a^m}] \ket{\Psi_{\rm R}}\,, 
\end{equation}
\begin{equation}\label{eq:eom_tauia_TDOCCT1}
    \begin{aligned}
&i(\delta_{b^m}^{a^m}D_{i^m}^{j^m}-D_{b^m}^{a^m}\delta_{i^m}^{j^m})\dot{\tau}_{j^m}^{b^m}  = \\
&B_{i^m}^{a^m} - i(\delta_{b^m}^{a^m}D_{i^m}^{j^m}-D_{b^m}^{a^m}\delta_{i^m}^{j^m})X_{j^m}^{b^m} -i\sum_n\frac{1}{2}A^{a^mj^n}_{i^mb^n}  (X^{b^n}_{j^n})^*\,. 
    \end{aligned}
\end{equation}
Interestingly, TD-OCCT1 is mathematically equivalent to spilt OATDCC (sOATDCC, or equivalently, sTDMVCC)\cite{hojlund2024bivariational}, which uses orthogonal virtual orbitals and bi-orthogonal active orbitals without single CC amplitudes and is reported in the vibrational systems and electron dynamics recently. The biorthogonal orbitlas in sOATDCC have the same role as $\hat{T}_1$ in the TD-OCCT1, however, they will give different $X_{i^m}^{j^m}$ and $X_{a^m}^{b^m}$ constraints, which might give different numerical performances. For more details, see Appendix.~\ref{Sec.sOATDCC}. When virtual orbitals do not exist, TD-OCCT1 and sOATDCC are equivalent to TDMVCC\cite{madsen2020time} and OATDCC\cite{Kvaal:2012} (NOCC)\cite{pedersen2001gauge} for vibrational systems and electron dynamics, respectively.

\subsubsection{TD-BCC}

Setting ${\tau}^{a^m}_{i^m}\equiv 0$ in Eq.~(\ref{eq:eom_t1},\ref{eq:eom_Xia}),
\begin{align}\label{eq:eom_Xia_TDBCC}
 iX_{i^m}^{a^m} + \sum_ni&\langle\Phi_{i^m}^{a^m}| e^{-\hat{T}} \hat{E}^{j^n}_{b^n}|\Psi_{\rm R}\rangle X^{j^n}_{b^n}  = \langle\Phi_{i^m}^{a^m}| e^{-\hat{T}} \hat{H}|\Psi_{\rm R}\rangle\,,
\end{align}
\begin{align}\label{eq:eom_lambdaai_TDBCC}
&i\frac{1}{2}\left\{\sum_n\langle\Phi_{j^n}^{b^n}| e^{-\hat{T}} \hat{E}^{i^m}_{a^m}|\Psi_{\rm R}\rangle\dot{\lambda}^{j^n}_{b^n} + (\dot{\lambda}^{i^m}_{a^m})^* \right\} = \nonumber \\
&B_{i^m}^{a^m}-i(\delta_{b^n}^{a^m}D_{i^m}^{j^n}-D_{b^n}^{a^m}\delta_{i^m}^{j^n})X_{j^n}^{b^n} -i\sum_n\frac{1}{2}A^{a^mj^n}_{i^mb^n}  (X^{b^n}_{j^n})^*  \,.  
\end{align}

\subsubsection{TD-OCCX0}
TD-OCCX0 can be considered as the most straightforward extension of TD-CC when virtual orbitals are considered, due to the similarity of the parameterization of the CC amplitudes parts and the absence of particle-hole orbital rotations. The EoMs of the single amplitudes of TD-OCCX0 are given by Eq.~(\ref{eq:eom_t1},\ref{eq:eom_l1}). The nonvariance of particle-hole orbital rotations makes the method not fully variational. Therefore, the Ehrenfest theorem should be modified\cite{Sato:2016} even when frozen-cores are excluded,  
\begin{equation}\label{eq:mEhren}
\begin{aligned}
         2i\frac{d}{dt}\braket{\hat{O}} &= \braket{\Psi_{\rm L}|[\hat{O},\hat{H}]|\Psi_{\rm R}} + \braket{\Psi_{\rm R}|[\hat{O},\hat{H}]|\Psi_{\rm L}} \\
        &+ \sum_m O_{t^m}^{u^m}\{\braket{\Psi_{\rm L}|(\hat{H}-i\hat{X})e^{\hat{T}}\bar{\hat{\Pi}}e^{-\hat{T}}\hat{E}_{t^m}^{u^m}|\Psi_{\rm R}}  \\
        &-\braket{\Psi_{\rm L}|\hat{E}_{t^m}^{u^m}e^{\hat{T}}\bar{\hat{\Pi}}e^{-\hat{T}}(\hat{H}-i\hat{X})|\Psi_{\rm R}} \\
        &+ \braket{\Psi_{\rm R}|(\hat{H}-i\hat{X})e^{\hat{T}}\bar{\hat{\Pi}}e^{-\hat{T}}\hat{E}_{t^m}^{u^m}|\Psi_{\rm L}}  \\
        &-\braket{\Psi_{\rm R}|\hat{E}_{t^m}^{u^m}e^{\hat{T}}\bar{\hat{\Pi}}e^{-\hat{T}}(\hat{H}-i\hat{X})|\Psi_{\rm L}} \}\,,
\end{aligned}
\end{equation}
where $\bar{\hat{\Pi}}=\hat{I}-\hat{\Pi}$. Similarly, TD-OCCH is not fully variational either when TD-OCCX0-type parameterization is considered, and its modified Ehrenfest theorem (as well as frozen-cores including scenario) will be presented in the Appendix.~\ref{Sec.TD-OCCH}. As a comparison, Ehrenfest theorem holds for other three methods since they are fully variational. 

When frozen orbitals are included, the modified Ehrenfest theorems\cite{Sato:2016} are necessary for all five methods: two additional terms that replace $t^m$, $u^m$ with $i^{\prime\prime m}$, $\mu^m$ and $\mu^m$, $i^{\prime\prime m}$ in the last term of Eq.~(\ref{eq:mEhren}) should be added in case of TD-OCCX0, and replace $t^m$, $u^m$ with $i^{\prime\prime m}$, $\mu^m$ and $\mu^m$, $i^{\prime\prime m}$ in the last term of Eq.~(\ref{eq:mEhren}) in case of TD-BCC, TD-OCC, and TD-OCCT1. The modified Ehrenfest theorem of TD-OCCH with frozen-cores will be reported in the Appendix.~\ref{Sec.TD-OCCH}.

\subsection{Discussion}


One of the major problems of TD-CC is that the growth of CC amplitudes could cause numerical instability\cite{pedersen2019symplectic,kristiansen2020numerical}, which can be mitigated by rotations of reference determinants that could compensate for excitations and reduce the amplitude norms $\tau$ and $\lambda$ in all five methods considered in this article. However, one must be careful when using TD-OCCX0 since such a compensation might not be huge when virtual orbital subspaces are relatively small due to the absence of particle-hole orbital rotations. 

Although the absence of particle-hole orbital rotations in TD-OCCX0 poses inconveniences for observable evaluations and numerical stability, it does save computing resources, especially for vibrational systems. Inverse matrix operations required to solve $X_{i^m}^{a^m}$ in other methods are unnecessary for TD-OCCX0. Comparable methods in this regard are TD-OCCT1, in which $X_{i^m}^{a^m}$ [Eq.~(\ref{eq:eom_Xia_TDOCCT1})] and $\tau_{i^m}^{a^m}$ [Eq.~(\ref{eq:eom_tauia_TDOCCT1})] of different modes do not couple with each other, and TD-OCCH with few fermion kinds using BCC and OCC types parameterization. For the details, see the Appendix.~\ref{Sec.TD-OCCH}. In both two methods, only several small-dimensional matrix inverse operations are required, which is much cheaper than TD-OCC [Eq.~(\ref{eq:eom_Xia_TDOCC})] and TD-BCC [Eq.~(\ref{eq:eom_Xia_TDBCC},\ref{eq:eom_lambdaai_TDBCC})] cases, requiring huge-dimensional matrix inverse operations. However, when strong external fields exist, the advantage of TD-OCCX0 and TD-OCCH can be dismissed due to the unfavorable resource cost of the evaluations of position matrix elements\cite{Sato:2016,sato2018gauge}, which will be discussed in Sec.~\ref{Sec.electron} and Appendix.~\ref{Sec.TD-OCCH}.

\section{Applications\label{sec:application}}
In this section, we will discuss three major applications of our methods that cover most of chemical systems: electron dynamics, nuclei dynamics, and nuclei-electron coupling dynamics. 

\subsection{Electron dynamics in a strong laser field}\label{Sec.electron}

For a many-electron system in external semi-classical electromagnetic fields, the Hamiltonian in the second quantization can be expressed as
\begin{eqnarray}\label{eq:ham}
&&\quad\quad\quad\quad  
\hat{H}(t)=h^\mu_\nu(t)\hat{E}^\mu_\nu+\frac{1}{2}g^{\mu\gamma}_{\nu\delta}\hat{E}^{\mu\gamma}_{\nu\delta},
 \\
&&\quad\quad\quad h^\mu_\nu(t)=\int dx\psi^*_\mu [h_0+V_{\rm ext}(t)]\psi_\nu, \\
&&g^{\mu\gamma}_{\nu\delta}=\int\int dx_1 dx_2\psi^*_\mu(x_1)\psi^*_\gamma(x_2) r^{-1}_{12}\psi_\nu(x_1)\psi_\delta(x_2),
\end{eqnarray}
where $x\!=\!(\pmb{r},\sigma)$ labels spatial-spin coordinate. Here, we drop off the kind lable for the simplicity (although one can treat different spin component as different kinds and use group excitation truncation scheme to save the computational resources). $V_{\rm ext}$ is the electron-laser interaction term. $h_0$ is the one-electron field free Hamiltonian that contains the kinetic term and Coulomb's terms induced by fixed nuclei. Under the framework of the electric dipole approximation, $V_{\rm ext}=\pmb{E}(t)\cdot\pmb{r}$ in the length gauge (LG) or $V_{\rm ext}=-i\pmb{A}(t)\cdot\pmb{\nabla}_{\pmb{r}}$ in the velocity gauge (VG), with $\pmb{E}(t)$ and $\pmb{A}(t)=-\int^t dt^\prime \pmb{E}(t^\prime)$ being the electric field and vector potential, respectively. In principle, two gauges are equivalent in the basis-set limit (basis numbers tend to the infinity), but inequivalent in practical numerical simulations due to different convergence speeds of basis-sets. Empirically, VG is more suitable for ultra strong fields scenario while LG is more suitable for other cases\cite{han2010comparison}.

When methods are fully variational, the gauge invariance of variational methods are satisfied automatically. Otherwise, the non-variance part of orbital rotations should be adjusted to ensure the gauge invariance. For the methods considered in this article, FC-part of $\hat{X}$ need to be selected as
\begin{equation}
    X_\mu^{i^{\prime\prime }} = i \pmb{E}(t)\cdot\braket{\psi_{i^{\prime\prime }}|\pmb{r}|\psi_{\mu}}\,,
\end{equation}
and the particle-hole orbital rotations of TD-OCCX0 should be
\begin{equation}
    X_{i}^{a} = i \pmb{E}(t)\cdot\braket{\psi_{a}|\pmb{r}|\psi_{i}}\,.\label{eq.VGLG}
\end{equation}
in the VG\cite{Sato:2016}. These adjustments are also necessary in the many-kind fermion systems, and the kind label should be explicitly included. Specifically, Eq.~(\ref{eq.VGLG}) should be applied in all kinds with TD-OCCX0 type parameterization in TD-OCCH method, see also Appendix.~\ref{Sec.TD-OCCH}.

\subsection{Vibrational systems}\label{Sec.Vibration}

The application of methods to vibrational problems is straightforward. A vibrational system with $m$ vibrational modes is isomorphic to a $m$-specie fermion system with only one fermion occupation in each specie. We point out that quantum photons, which are essential for molecules in cavities, can also be described in the same framework. 

Previous works of TDVCC families\cite{hojlund2024time,hojlund2024bivariational,hojlund2022bivariational,hansen2019time,madsen2020time} focus mainly on the $[k]$-truncation scheme, which is sufficient for ordinary systems. Nonetheless, for the systems with double well potentials and vibronic coupling systems (see next subsection), weighted excitation truncation scheme or group excitation truncation scheme is necessary for the faithful simulations since the degree of freedom (DOF) with double well potentials has more contribution than other DOFs to the dynamics. A practical strategy of weighted excitation truncation scheme for such a system is to set the DOF with double well potentials as 0 and other DOFs as 1. Currently, special truncation schemes have been successfully implemented in TDVCC\cite{madsen2020general} and TDMVCC\cite{jensen2023efficient,TDH-TDMVCC}, but lacking theoretical discussion on orbital rotation invariance. The proof of the intragroup rotations of particle and hole orbitals in the Appendix.~\ref{Sec.Redundancy} also holds for the biorthogonal orbitals scenario. Therefore, we show that weighted excitation truncation scheme can be safely used in TDVCC families without compromising the intragroup rotations of particle and hole orbitals, which benefits future numerical implementations. Despite existing TDVCC families and weighted excitation truncation schemes, our work provides new ansatz, TD-OCCT1, TD-BCC, TD-OCCX0, and TD-OCCH, and the group excitation truncation scheme as new candidates for future implementations. Additionally, although TD-OCCT1 and sTDMVCC are mathematically equivalent, different constraints $X_{i^m}^{j^m}$ and $X_{a^m}^{b^m}$, as well as using orthogonal or nonorthogonal orbitals, might give difference performances in practical implementations. Therefore, they should be benchmarked systematically.

\subsection{Vibronic coupling systems}

There are two ways to simulate vibronic coupling systems. Traditionally, the electronic DOF is described as a $S$-level system, thus a vibronic coupling system is isomorphic to a vibrational system. The major differences are: 1. almost always all $S$ orbitals of the electronic DOF are active in the wavefunction ansatz (no virtual orbitals); 2. vibrational contributions to the dynamics associated with different electronic levels are comparable\cite{worth2004beyond,kouppel1984multimode,bao2024time}. These make TD-OCCX0 and TD-OCC unsuitable for the vibronic coupling simulations, and one should also avoid using OCCX0 and OCC types parameterization for the electronic DOF in TD-OCCH. Additionally, the same weighted excitation truncation scheme as systems with the double well potentials (proposed in the last subsection) should be used.

In recent years, treating electronic and nuclear DOF on an equal footing has attracted increasing attention due to advancements in theory\cite{sibaev2020molecular,muolo2020nuclear,sasmal2020non,sasmal2022sum,sasmal2024compact}. The five CC ansatz and two special truncation schemes proposed in this article can be regarded as the prototype of future advanced CC simulations.

\section{Summary\label{sec:sum}}
We have successfully formulated five new orbital-optimized time-dependent coupled-cluster methods, which are distinguished by whether single amplitudes, $\tau_{i^m}^{a^m}$, $\lambda_{a^m}^{i^m}$ and particle-hole orbital rotations $X_{i^m}^{a^m}$ are constraints, for arbitrary fermion mixtures. When one of $\tau_{i^m}^{a^m}$, $\lambda_{a^m}^{i^m}$, and $X_{i^m}^{a^m}$ is a constraint for each fermion kind $m$, the method can converge to the CASSCF. Two advanced truncation schemes of higher-order amplitudes which maintain the intragroup rotation invariance are introduced. The applications to electronic dynamics, vibrational dynamics, and non-adiabatic dynamics are also discussed. Our methods are more compact CC-parameterization alternatives of the CI-parameterization of TD-MCSCF method, and would shed light on the high-accuracy numerical simulations of large systems. The numerical implementations on the electronic dynamics will be presented in a forthcoming article.

\begin{acknowledgments}
This research was supported in part by a Grant-in-Aid for
Scientific Research Grant Number JP22H05025 from the Ministry of Education, Culture, Sports, Science and Technology (MEXT) of Japan. This research was also partially supported by MEXT
Quantum Leap Flagship Program (MEXT Q-LEAP) Grant
Number JPMXS0118067246 and JST COI-NEXT Grant
Number JPMJPF2221. This work is partly supported by IBM-UTokyo lab.
\end{acknowledgments}

\section*{AUTHOR DECLARATIONS}
\subsection*{Conflict of Interest}
The authors have no conflicts to disclose.
\subsection*{Author Contributions}
\textbf{Haifeng Lang}: Conceptualization (equal); Methodology (lead); Investigation (lead); Writing – original draft (lead); Writing –
review \& editing (equal). \textbf{Takeshi Sato}: Conceptualization (equal); Methodology (supporting); Funding acquisition
(lead);  Writing –review \&  editing (equal).

\section*{DATA AVAILABILITY}
The data that supports the findings of this study are available within the article.

\appendix

\section{Redundancy transfer of $X_{i^m}^{a^m}$, $\tau_{i^m}^{a^m}$ and ${\lambda}^{i^m}_{a^m}$ in the full CC limit}

In this section, we will demonstrate that the redundancy of $X_{i^m}^{a^m}$ can be transferred to $\lambda^{i^m}_{a^m}$ or $\tau_{i^m}^{a^m}$, and the preparation of the exact wavefunction without $\lambda^{i^m}_{a^m}$ or $\tau_{i^m}^{a^m}$ is possible in the full CC limit. The analysis on $\lambda^{i^m}_{a^m}$ and $\tau_{i^m}^{a^m}$ are extremely similar, for convenience, we will consider the case of all $\lambda^{i^m}_{a^m}$ redundancy first and briefly discuss the most general scenario.

Mathematically, the redundancy of $X_{i^m}^{a^m}$ means that the coefficient of $\Delta_{i^m}^{a^m}$ is identically zero, i.e,  Eq.~(\ref{eq:eom_Xia}) is identiy regardless the choice of $X_{i^m}^{a^m}$. This can be rigorously proved using the same method in the next section. 
Now we assume that the time-dependent trajectories of $\lambda^{i^m}_{a^m}$ are arbitrarily given. Eq.~(\ref{eq:eom_l1}) becomes equations of $X_{i^m}^{a^m}$. If the solution of Eq.~(\ref{eq:eom_l1}) exists and is unique, Eq.~(\ref{eq:eom_t1},\ref{eq:eom_Xia}) should hold due to the redundancy of $X_{i^m}^{a^m}$. Further, if one treats Eq.~(\ref{eq:eom_Xia}) as equations of $\dot{\tau}_{i^m}^{a^m}$ and assumes the solution exists and is unique, the redundancy of $X_{i^m}^{a^m}$ has been successfully transferred to $\lambda^{i^m}_{a^m}$. In general, it is very hard to prove the existence and uniqueness of the solution even without constraint $X_{a^m}^{b^m} = X_{i^m}^{j^m} \equiv 0$, but one can expect that the statement is true, at least when the state is dominantly of single reference character\cite{myhre2018demonstrating}.

To completely exclude $\lambda^{i^m}_{a^m}$ in the time-dependent full CC expansion, one needs to prepare the initial state with the parameterization $\lambda^{i^m}_{a^m}(0)=0$ and use the constraint $\dot{\lambda}^{i^m}_{a^m}\equiv0$. Here, we provide two possible methods to convert the parameterization of an arbitrary state $\ket{\Psi}$ into the vanishing $\lambda^{i^m}_{a^m}$ form. The first is to prepare the initial state as an arbitrary Slater determinant $ \ket{\Phi_0}$, and let it evolve from $t = 0$ to $t = \pi/2$ under the Hamiltonian $\hat{H}=-i(\ket{\Phi_0}\bra{\Psi} - \ket{\Psi}\bra{\Phi_0})$ with $\dot{\lambda}^{i^m}_{a^m}\equiv0$. The second is to let the state $\ket{\Psi}$ in an arbitrary parameterization form evolve from $t = 0$ to $t = 1$ under the Hamiltonian $\hat{H}=0$ with $\dot{\lambda}^{i^m}_{a^m}\equiv -\lambda^{i^m}_{a^m}(0)$.

Now we are in the position to give the most general redundancy analysis of parameters. Assuming one of $X_{i^m}^{a^m}$, $\lambda^{i^m}_{a^m}$, and $\tau_{i^m}^{a^m}$ is pre-determined for each kind of fermion, equations corresponding to the variation coefficients in $\delta S$ should be removed in coupled equations Eq.~(\ref{eq:eom_t1},\ref{eq:eom_l1},\ref{eq:eom_Xia}). If the solution of remaining equations exists and is unique, the removing equations are identities and the redundancy of parameters have been proved. To exclude these redundant parameters in the time-dependent full CC expansion, one may follow the same procedures as for the $\lambda^{i^m}_{a^m}$-redundancy scenario described above, with the replacements with corresponding pre-determined variables (constraints).

\section{Redundancy proof of particle-particle and hole-hole orbital rotations}\label{Sec.Redundancy}

We will explicitly show the redundancy of $X_{i^m}^{j^m}$ and $X_{a^m}^{b^m}$ from $\delta S$. For the scenarios $\mu^m,\nu^m \in \{i^m\}$, and $\mu^m,\nu^m \in \{a^m\}$,
    \begin{align}
        &\langle\Psi_{\rm L}|[\hat{H}-i(\dtp\hat{T})-i\hat{X},\hat{E}^{\mu^m}_{\nu^m}]|\Psi_{\rm R}\rangle \nonumber\\
+&i\langle\Phi|(\dtp\hat{\Lambda})e^{-\hat{T}}\hat{E}^{\mu^m}_{\nu^m}|\Psi_{\rm R}\rangle \nonumber\\
=& \langle\Psi_{\rm L}|[ \hat{H} - i\hat{X} ,\hat{E}^{\mu^m}_{\nu^m}]|\Psi_{\rm R}\rangle\nonumber \\
-&i\dot{\tau}_{\mathring{I}}^{\mathring{A}}\langle\Psi_{\rm L}|[\hat{E}_{\mathring{I}}^{\mathring{A}},\hat{E}^{\mu^m}_{\nu^m}]|\Psi_{\rm R}\rangle
+i\dot{\lambda}^{\mathring{I}}_{\mathring{A}}\langle\Phi_{\mathring{I}}^{\mathring{A}}|e^{-\hat{T}}\hat{E}^{\mu^m}_{\nu^m}|\Psi_{\rm R}\rangle \label{eq:General-exc}\\
=&\braket{\Psi_{\rm L}|\hat{E}_{\mathring{I}}^{\mathring{A}}(\hat{H} - i\hat{X} )|\Psi_{\rm R}}\langle\Phi_{\mathring{I}}^{\mathring{A}}|e^{-\hat{T}}\hat{E}^{\mu^m}_{\nu^m}|\Psi_{\rm R}\rangle \nonumber\\
-&\braket{\Phi_{\mathring{I}}^{\mathring{A}}|e^{-\hat{T}}(\hat{H} - i\hat{X} )|\Psi_{\rm R}}\langle\Psi_{\rm L}|\hat{E}_{\mathring{I}}^{\mathring{A}}\hat{E}^{\mu^m}_{\nu^m}|\Psi_{\rm R}\rangle \label{eq:BeforeRegroup}
\,, 
    \end{align}
Here, time derivatives of all single amplitudes are included in Eq.~(\ref{eq:General-exc}). 
Converting Eq.~(\ref{eq:General-exc}) to Eq.~(\ref{eq:BeforeRegroup}), one needs to replace all EoMs of CC amplitudes with Eq.~(\ref{eq:eom_t0},\ref{eq:eom_t1},\ref{eq:eom_l0},\ref{eq:eom_l1}) and use Eq.~(\ref{eq:criteria}). The equations are straightforward for TD-OCCX0, and we will explain more on other methods.  We first consider TD-OCCT1. $\lambda_{a^m}^{i^m}$ in the ansatz should be understood as a constraint $\lambda_{a^m}^{i^m}\equiv 0$. Therefore, Eq.~(\ref{eq:General-exc}) holds. Noticing that the multiplier of $\dot{\tau}^{a^m}_{i^m}$ in Eq.~(\ref{eq:General-exc}) vanishes, one can safely replace $\dot{\tau}^{a^m}_{i^m}$ with Eq.~(\ref{eq:eom_t1}). Additionally, Eq.~(\ref{eq:eom_l1}) also ensures $\dot{\lambda}_{a^m}^{i^m}\equiv 0$, and we have finished the proof of equations for TD-OCCT1. For the discussion on TD-BCC, one just needs to swap $\lambda_{a^m}^{i^m}$ and $\tau^{a^m}_{i^m}$ in the proof of TD-OCCT1. For the TD-OCC method, both $\lambda_{a^m}^{i^m}$ and $\tau^{a^m}_{i^m}$ should be regarded as constraints and their multipliers in Eq.~(\ref{eq:General-exc}) are always zero, and thus equations hold. Additionally, equations of TD-OCCH automatically hold when equations of four other methods hold.

Notice that $\langle\Phi^{\mathring{B}}_{\mathring{J}}|\hat{E}_{\mathring{I}}^{\mathring{A}}$ is not zero iff $\mathring{A}$ and $\mathring{I}$ are substrings of $\mathring{B}$ and $\mathring{J}$, respectively. For convenience, we use ${\mathring{B}} - {\mathring{A}}$ to express the absolute complementary string of $\mathring{A}$ with respect to $\mathring{B}$. Expanding $\bra{\Psi_{\rm L}}$ in Eq.~(\ref{eq:BeforeRegroup}) as $\lambda_{\mathring{B}}^{\mathring{J}}\bra{\Phi^{\mathring{B}}_{\mathring{J}}}$, Eq.~(\ref{eq:BeforeRegroup}) can be converted to 
\begin{equation}
    \begin{aligned}
        \lambda_{\mathring{B}}^{\mathring{J}}&\big( \bra{\Phi^{\mathring{B} - \mathring{A}}_{\mathring{J} - \mathring{I}}}(\hat{H} - i\hat{X})\ket{\Psi_{\rm R}}{\bra{\Phi^{\mathring{A}}_{\mathring{I}}}}e^{-\hat{T}}\hat{E}_{\nu^m}^{\mu^m}\ket{\Psi_{\rm R}} \\
        &- {\bra{\Phi^{\mathring{A}}_{\mathring{I}}}}(\hat{H} - i\hat{X})\ket{\Psi_{\rm R}}\bra{\Phi^{\mathring{B} - \mathring{A}}_{\mathring{J} - \mathring{I}}}e^{-\hat{T}}\hat{E}_{\nu^m}^{\mu^m}\ket{\Psi_{\rm R}}\big)\,.
    \end{aligned}
\end{equation}
Resuming the second term by swapping ${\mathring{B}} - {\mathring{A}}$, ${\mathring{J}} - {\mathring{I}}$ and ${\mathring{A}}$, ${\mathring{I}} $, one can immediately obtain that Eq.~(\ref{eq:BeforeRegroup}) is zero, which completes the proof that $X_{i^m}^{j^n}$ and $X_{a^m}^{b^n}$ are redundant. 

Following the same discussion, one can also explicitly show that $\hat{X}$ is redundant for TD-ooCC with $\hat{T}_1$ and $\hat{\Lambda}_1$ in the CASSCF limit. In this case, it is easy to check that Eq.~(\ref{eq:General-exc}) and Eq.~(\ref{eq:BeforeRegroup}) also hold, and applying the same resummation argument as the proof of redundancy of $X_{i^m}^{j^m}$ and $X^{b^m}_{a^m}$, the proof of redundancy of $X_{i^m}^{a^m}$ can be obtained.

\section{OATDCC and OATDCC with split orthogonal and biorthogonal orbitals}\label{Sec.sOATDCC}

Before presenting the equivalence proof of sTDMVCC\cite{hojlund2024bivariational} (sOATDCC) and TD-OCCT1, we will briefly review TDVCC families, including TDVCC\cite{hansen2019time}, TDMVCC, restricted polar TDMVCC (rpTDMVCC)\cite{hojlund2022bivariational}, orthogonal TDMVCC (oTDMVCC)\cite{hojlund2024time}, and sTDMVCC\cite{hojlund2024bivariational}. First, we will focus on the EoMs and parameterization of singlel amplitudes of CC coefficients and orbitals. In TDVCC, all orbitals are active, and TD-OCCX0 is equivalent to TDVCC in this case. The fast growth of norms of CC amplitudes arisen from the absence of orbital rotations makes applications of TDVCC limited. It is the vibrational analog of TD-CC. To resolve this problem, orbital rotations and virtual orbitals are introduced. All other four methods, TDMVCC, rpTDMVCC, sTDMVCC and oTDMVCC have formally same (but not equivalent due to orthogonal/biorthogonal orbitals) CC part parameterizations and EoMs, and their major differences are the parameterization of orbitals and their EoMs. oTDMVCC is completely identical to TD-OCC\cite{sato2018gauge} of this article, which uses orthogonal orbitals but suffers from the huge matrix inversion and nonconvergence to CASSCF. Instead of using orthogonal orbitals, TDMVCC, which is the vibrational analog of OATDCC, solves the matrix inversion and convergence problems by using biorthogonal orbitals, but the non-orthogonality causes numerical instabilities in long time simulations. One can force active orbitals lie in the conjugate ket and bra spaces to mitigate instabilities, which is implemented by using orthogonal virtual orbitals and biorthogonal active orbitals in rpTDMVCC and sTDMVCC. EoMs of active-virtual orbital rotations are determined by the variational principle and projection in sTDMVCC and rpTDMVCC, respectively, while EoMs of CC amplitudes and particle-hole rotations of two methods are completely identical. In fact, the contribution of biorthogonal active orbitals is equivalent to the one-rank excitation in TD-OCCT1, and the TDVP ensures the equivalence of TD-OCCT1 and sTDMVCC.

For convenience, we only consider models without dynamical-core in this Appendix. For models with dynamical-core, the treatments of dynamical-core orbitals are almost identical to the virtual orbitals except that they should be evolved explicitly. We will first review the standard OATDCC, which uses birothogonal bra and ket orbitals rather than orthogonal orbitals in TD-OCC. The highly flexible variational space choices could cause serious numerical instabilities arisen from the different spaces spanned by ket and bra active orbitals in practical implementations\cite{hojlund2022bivariational}. Next, we will review sOATDCC, i.e., OTADCC with virtual orbitals of ket and bra are orthogonal in the Lagrangian and active orbitals are biorthogonal. Such a parameterization forces the spaces spanned by active orbitals of ket and bra are identical, which eliminates the numerical instability. The modifications only appear in the intergroup rotation between active orbitals and other types of orbitals. OATDCC and sOATDCC reviewed in this Appendix are describing fermion mixtures with arbitrary occupations, which are more general than the references results. When sOATDCC is applied to the vibrational problem, it is the sTDMVCC and inherits the advantage of TDMVCC\cite{madsen2020time,hojlund2024time}: EoMs of different vibrational modes are separate and converge to the CASSCF. In contrast to rpTDMVCC\cite{hojlund2022bivariational}, this approach is fully variational. We further prove that this approach is equivalent to TD-OCCT1.

We define $\ket{\phi_{\mu^m}}$ and $\bra{\Tilde{\phi}_{\mu^m}}$ as biorthogonal bases of OATDCC, and their corresponding creation (annihilation) operators are denoted as $\hat{a}_{\mu^m}^\dagger$ ($\hat{a}_{\mu^m}$) and $\Tilde{\hat{a}}_{\mu^m}^\dagger$ ($\Tilde{\hat{a}}_{\mu^m}$). They satisfy the following relations,
\begin{align}
    & \braket{\Tilde{\phi}_{\mu^m}|\phi_{\nu^m}} = \delta_{\mu^m\nu^m}\,,\\
    & \{\hat{a}_{\nu^m}^\dagger,\Tilde{\hat{a}}_{\mu^m}\} = \braket{\Tilde{\phi}_{\mu^m}|\phi_{\nu^m}} = \delta_{\mu^m\nu^m}\,.
\end{align}
In the standard OATDCC, $\hat{a}_{\mu^m}$ and $\Tilde{\hat{a}}_{\mu^m}^\dagger$ are never used, nonetheless, they play important roles in the sOATDCC. We also introduce the operator $\Tilde{\hat{E}}^{\mu_1^{m_1}\mu_2^{m_2}\mu_3^{m_3}\cdots}_{\nu_1^{m_1}\nu_2^{m_2}\nu_3^{m_3}\cdots}=\hat{a}^\dagger_{\mu_1^{m_1}}\hat{a}^\dagger_{\mu_2^{m_2}}\hat{a}^\dagger_{\mu_3^{m_3}}\cdots\Tilde{\hat{a}}_{\nu_3^{m_3}}\Tilde{\hat{a}}_{\nu_2^{m_2}}\Tilde{\hat{a}}_{\nu_1^{m_1}}$, and all other quantities in OATDCC are denoted as corresponding quantities in TD-OCC with tildes, similarly. $m_1,m_2,\cdots,m_3$ can be the same index of fermion kind. The ket and bra reference wavefunctions of OATDCC are tensor products Slater determinants of $\ket{\phi_{\mu^m}}$ and $\bra{\Tilde{\phi}_{\mu^m}}$ for all $m$, denoted as $\ket{{\Phi}}$ and $\bra{\Tilde{\Phi}}$, respectively. Wavefunction and their time derivatives and variations as well as the Lagrangian in OATDCC are formally identical to the TD-OCC correspondence with tildes replacements. Due to the  loss of orthogonality, $(\Tilde{\hat{E}}_I^A)^\dagger \neq \Tilde{\hat{E}}_A^I$, $(\Tilde{X}^{\mu^m}_{\nu^m})^* \neq \Tilde{X}_{\mu^m}^{\nu^m}$, and  $(\Tilde{\Delta}^{\mu^m}_{\nu^m})^* \neq \Tilde{\Delta}_{\mu^m}^{\nu^m}$ in general. Also, the anti-Hermicity of $\Tilde{\hat{X}}$ and $\Tilde{\hat{\Delta}}$ cannot be expected. Therefore, all variables and their complex conjugate are independent quantities in OATDCC. The action of OATDCC is a complex functional
\begin{eqnarray}
\Tilde{S} &=& \label{eq:actionOATD}
 \int_{t_0}^{t_1}dt  \Tilde{L} \,,
\end{eqnarray}
and require the action to be stationary
\begin{eqnarray}
{\color{black}}\delta \Tilde{S} &=& \label{eq:vars0b}
\langle\delta\Tilde{\Psi}_{\rm L}|\hat{H}|\Tilde{\Psi}_{\rm R}\rangle
+\langle\Tilde{\Psi}_{\rm L}|\hat{H}|\delta\Tilde{\Psi}_{\rm R}\rangle \nonumber \\
&-&i\langle\delta\Tilde{\Psi}_{\rm L}|\dot{\Tilde{\Psi}}_{\rm R}\rangle
+i\langle\dot{\Tilde{\Psi}}_{\rm L}|\delta\Tilde{\Psi}_{\rm R}\rangle  = 0\,.
\end{eqnarray}
The non-existence of complex conjugate variables in OATDCC action makes the action complex analytic, and ensures the existence of the solution. The complex even dimension of the variational manifold of OATDCC ensures uniqueness of the solution. OTADCC converges to CASSCF when all excitation and deexcitation operators except for the single excitation and deexcitation ones are included. We also point out that the real action choice will lead the same solutions due to the independence of all variables and their complex conjugates.

Working out the variations explicitly
\begin{widetext}
{\color{black}
\begin{eqnarray} 
\delta \Tilde{S} 
&=&
\delta\Tilde{\tau}^A_I\left\{
\langle\Tilde{\Psi}_{\rm L}|
[\hat{H}-i\Tilde{\hat{X}},\Tilde{\hat{E}}^A_I]|\Tilde{\Psi}_{\rm R}\rangle
+i\dot{\Tilde{\lambda}}^I_A\right\} 
+\delta\Tilde{\lambda}^I_A\left\{
\langle\Tilde{\Phi}^A_I|e^{-\Tilde{\hat{T}}}
(\hat{H}-i\Tilde{\hat{X}})e^{\Tilde{\hat{T}}}|\Tilde{\Phi}\rangle
-i\dot{\Tilde{\tau}}^A_I\right\}  \nonumber \\
&+&
\Tilde{\Delta}^{\mu^m}_{\nu^m} \left\{
\langle\Tilde{\Psi}_{\rm L}|[\hat{H}-i(\dtp\Tilde{\hat{T}})-i\Tilde{\hat{X}},\Tilde{\hat{E}}^{\mu^m}_{\nu^m}]|\Tilde{\Psi}_{\rm R}\rangle
+i\langle\Tilde{\Phi}|(\dtp\Tilde{\hat{\Lambda}})e^{-\Tilde{\hat{T}}}\Tilde{\hat{E}}^{\mu^m}_{\nu^m}e^{\Tilde{\hat{T}}}|\Tilde{\Phi}\rangle
 \right\}\,,\label{eq:vars_orbOATD}
\end{eqnarray}
}
\end{widetext}
%
{\color{black}
EoMs of CC coefficients, 
\begin{eqnarray}
i\dot{\Tilde{\tau}}^A_I &=& \label{eq:eom_t}
\langle\Tilde{\Phi}_I^A| e^{-\Tilde{\hat{T}}} (\hat{H}-i\Tilde{\hat{X}})e^{\Tilde{\hat{T}}}|\Tilde{\Phi}\rangle\,, \\
-i\dot{\Tilde{\lambda}}^I_A &=& \label{eq:eom_l}
\langle\Tilde{\Psi}_{\rm L}|
[\hat{H}-i\Tilde{\hat{X}},\Tilde{\hat{E}}^A_I]|\Tilde{\Psi}_{\rm R}\rangle\,,
\end{eqnarray}
and EoMs of orbital rotations,
\begin{align}
        &\bra{\Tilde{\Psi}_{\rm L}}[\hat{H} - i\Tilde{\hat{X}},\Tilde{\hat{E}}^{\alpha^m}_{p^m}]\ket{\Tilde{\Psi}_{\rm R}} =0\,,\\
        &\bra{\Tilde{\Psi}_{\rm L}}[\hat{H} - i\Tilde{\hat{X}},\Tilde{\hat{E}}_{\alpha^m}^{p^m}]\ket{\Tilde{\Psi}_{\rm R}} =0\,,\\
        &\bra{\Tilde{\Psi}_{\rm L}}[\hat{H} - i\Tilde{\hat{X}},\Tilde{\hat{E}}_{i^m}^{a^m}]\ket{\Tilde{\Psi}_{\rm R}} =0\,,\\
        &\langle\Tilde{\Psi}_{\rm L}|[\hat{H}-i(\dtp\Tilde{\hat{T}})-i\Tilde{\hat{X}},\Tilde{\hat{E}}_{a^m}^{i^m}]|\Tilde{\Psi}_{\rm R}\rangle \nonumber \\
+&i\langle\Tilde{\Phi}|(\dtp\Tilde{\hat{\Lambda}})e^{-\Tilde{\hat{T}}}\Tilde{\hat{E}}_{a^m}^{i^m}e^{\Tilde{\hat{T}}}|\Tilde{\Phi}\rangle = 0\,,
\end{align}
can be obtained. Similar to the TD-OCC scenario, all intragroup rotations are redundant, thus should be pre-selected, for instance, setting all the redundant rotations as zero.

In principle, $\{\ket{\phi_{p^m}}\}$ and $\{\ket{\Tilde{\phi}_{p^m}}\}$ can span different spaces, which can make the propagation unstable. To resolve this problem, we adapt real action with the constraints that virtual orbitals are orthogonal,
\begin{eqnarray}
S_s &=& \label{eq:actionC}
 \frac{1}{2}\int_{t_0}^{t_1}dt \left( \Tilde{L} + \Tilde{L}^*  \right),
\end{eqnarray}
We still use the notations of OATDCC in sOATDCC when there is no ambiguity. In this case, corresponding ket and bra virtual orbitals are identical, $\ket{\phi_{\alpha^m}}^* = \bra{\Tilde{\phi}_{\alpha^m}}$, $\hat{a}_{\alpha^m} = \Tilde{\hat{a}}_{\alpha^m}$ and $\hat{a}_{\alpha^m}^\dagger = \Tilde{\hat{a}}_{\alpha^m}^\dagger$. Therefore, $\{\ket{\phi_{p^m}}\}$ and $\{\ket{\Tilde{\phi}_{p^m}}\}$ span the same space, and the following relations hold,
\begin{equation}
    \begin{aligned}
        \ket{\Tilde{\phi}_{p^m}} = \braket{\Tilde{\phi}_{q^m}|\Tilde{\phi}_{p^m}}\ket{\phi_{q^m}}\,, \quad\Tilde{\hat{a}}_{p^m}^\dagger = \braket{\Tilde{\phi}_{q^m}|\Tilde{\phi}_{p^m}} \hat{a}_{q^m}^\dagger \,,\\
        \bra{\phi_{p^m}} = \bra{\Tilde{\phi}_{q^m}}\braket{\phi_{p^m}|\phi_{q^m}} \,, \quad \hat{a}_{p^m} = \braket{\phi_{p^m}|\phi_{q^m}}\Tilde{\hat{a}}_{q^m} \,.
    \end{aligned}
\end{equation}
The constraints also force $\ket{\delta\phi_{\alpha^m}} \equiv \ket{\delta\Tilde{\phi}_{\alpha^m}} $ and $\ket{\dot{\phi}_{\alpha^m}} \equiv \ket{\dot{\Tilde{\phi}}_{\alpha^m}} $. Using above relations, one could find that $(\Tilde{X}^{\alpha^m}_{p^m})^*$ and $\Tilde{X}_{\alpha^m}^{p^m}$, as well as $(\Tilde{\Delta}^{\alpha^m}_{q^m})^*$ and $\Tilde{\Delta}_{\alpha^m}^{q^m}$ are not independent,
\begin{align}
        &(\Tilde{X}^{\alpha^m}_{p^m})^* = - \Tilde{X}_{\alpha^m}^{q^m}\braket{\phi_{p^m}|\phi_{q^m}}\,,\\
        &\Tilde{X}_{\alpha^m}^{p^m} = - (\Tilde{X}^{\alpha^m}_{q^m})^*\braket{\Tilde{\phi}_{p^m}|\Tilde{\phi}_{q^m}}\,,\\
         &(\Tilde{\Delta}^{\alpha^m}_{p^m})^* = - \Tilde{\Delta}_{\alpha^m}^{q^m}\braket{\phi_{p^m}|\phi_{q^m}}\,,\\
         &\Tilde{\Delta}_{\alpha^m}^{p^m} = - (\Tilde{\Delta}^{\alpha^m}_{q^m})^*\braket{\Tilde{\phi}_{p^m}|\Tilde{\phi}_{q^m}}\,,\\
                &(\Tilde{X}^{\alpha^m}_{p^m}\Tilde{\hat{E}}^{\alpha^m}_{p^m})^\dagger =  -\Tilde{X}_{\alpha^m}^{p^m}\Tilde{\hat{E}}_{\alpha^m}^{p^m}\,,\\
                &(\Tilde{\Delta}^{\alpha^m}_{p^m}\Tilde{\hat{E}}^{\alpha^m}_{p^m})^\dagger =  -\Tilde{\Delta}_{\alpha^m}^{p^m}\Tilde{\hat{E}}_{\alpha^m}^{p^m}\,.
\end{align}

Therefore, the explicit form of the variation with independent variational variables is
\begin{widetext}
{\color{black}
\begin{eqnarray} 
2\delta {S}_s 
&=&
\delta\Tilde{\tau}^A_I\left\{
\langle\Tilde{\Psi}_{\rm L}|
[\hat{H}-i\Tilde{\hat{X}},\Tilde{\hat{E}}^A_I]|\Tilde{\Psi}_{\rm R}\rangle
+i\dot{\Tilde{\lambda}}^I_A\right\} 
+\delta\Tilde{\lambda}^I_A\left\{
\langle\Tilde{\Phi}^A_I|e^{-\Tilde{\hat{T}}}
(\hat{H}-i\Tilde{\hat{X}})e^{\Tilde{\hat{T}}}|\Tilde{\Phi}\rangle
-i\dot{\Tilde{\tau}}^A_I\right\}  + c.c. \nonumber \\
&+&
\Tilde{\Delta}^{p^m}_{q^m} \left\{
\langle\Tilde{\Psi}_{\rm L}|[\hat{H}-i(\dtp\Tilde{\hat{T}})-i\Tilde{\hat{X}},\Tilde{\hat{E}}^{p^m}_{q^m}]|\Tilde{\Psi}_{\rm R}\rangle
+i\langle\Tilde{\Phi}|(\dtp\Tilde{\hat{\Lambda}})e^{-\Tilde{\hat{T}}}\Tilde{\hat{E}}^{p^m}_{q^m}e^{\Tilde{\hat{T}}}|\Tilde{\Phi}\rangle
 \right\}\,+ c.c. \nonumber \\
 &+&
\Tilde{\Delta}^{p^m}_{\alpha^m} \left\{
\langle\Tilde{\Psi}_{\rm L}|[\hat{H}-i\Tilde{\hat{X}},\Tilde{\hat{E}}^{p^m}_{\alpha^m}]|\Tilde{\Psi}_{\rm R}\rangle - \braket{\phi_{q^m}|\phi_{p^m}}\langle\Tilde{\Psi}_{\rm L}|[\hat{H}-i\Tilde{\hat{X}},\Tilde{\hat{E}}_{q^m}^{\alpha^m}]|\Tilde{\Psi}_{\rm R}\rangle^*
 \right\} \nonumber \\
  &+&
\Tilde{\Delta}_{p^m}^{\alpha^m} \left\{
\langle\Tilde{\Psi}_{\rm L}|[\hat{H}-i\Tilde{\hat{X}},\Tilde{\hat{E}}_{p^m}^{\alpha^m}]|\Tilde{\Psi}_{\rm R}\rangle - \braket{\Tilde{\phi}_{p^m}|\Tilde{\phi}_{q^m}}\langle\Tilde{\Psi}_{\rm L}|[\hat{H}-i\Tilde{\hat{X}},\Tilde{\hat{E}}^{q^m}_{\alpha^m}]|\Tilde{\Psi}_{\rm R}\rangle^*
 \right\}\,.
 \label{eq:vars_orbcOATD}
\end{eqnarray}
}
\end{widetext}
Compared with Eq.~(\ref{eq:vars_orbOATD}), it is clear that the constraints do not change the EoMs of coefficients and intergroup orbital rotations between particle and hole orbitals. The variations with respect to $\Delta_{\alpha^m}^{p^m}$ give EoMs of intergroup rotation between active orbitals and virtual orbitals 

\begin{align}
    &\bra{\Tilde{\Psi}_{\rm L}}[\hat{H} - i\Tilde{\hat{X}},\Tilde{\hat{E}}_{\alpha^m}^{p^m}]\ket{\Tilde{\Psi}_{\rm R}} + \bra{\Tilde{\Psi}_{\rm R}}[\hat{H} - i\Tilde{\hat{X}},\Tilde{\hat{E}}_{\alpha^m}^{p^m}]\ket{\Tilde{\Psi}_{\rm L}} =0\,,\\
&\bra{\Tilde{\Psi}_{\rm L}}[\hat{H} - i\Tilde{\hat{X}},\Tilde{\hat{E}}^{\alpha^m}_{p^m}]\ket{\Tilde{\Psi}_{\rm R}} + \bra{\Tilde{\Psi}_{\rm R}}[\hat{H} - i\Tilde{\hat{X}},\Tilde{\hat{E}}^{\alpha^m}_{p^m}]\ket{\Tilde{\Psi}_{\rm L}} =0\,.
\end{align}

sOATDCC is equivalent to TD-OCCT1. To see this, one just needs to prove their manifolds are identical up to normalization factors, which corresponds to a point transformation in Lagrange Mechanics\cite{Herbert:1980}. More explicitly speaking, for any states $\ket{\Tilde{\Psi}_{\rm R}}$ and $\bra{\Tilde{\Psi}_{\rm L}}$ parameterized via the sOATDCC formalism, one can always find states parameterized via the TD-OCCT1 formalism $\ket{\Psi_{\rm R}}$ and $\bra{\Psi_{\rm L}}$ that satisfy $\ket{\Tilde{\Psi}_{\rm R}} = A\ket{\Psi_{\rm R}}$, and $\bra{\Tilde{\Psi}_{\rm L}} = A^{-1}\bra{\Psi_{\rm L}}$ and vice versa. Here, $A$ is a normalization factor and only gives a trivial total derivative in Lagrange. In both transformations, one can always choose $\ket{\psi_\alpha}\equiv\ket{\phi_\alpha}$ and they only appear in the time derivatives of wavefunctions. 

We will first consider converting TD-OCCT1 to sOATDCC. The transformation is given by
\begin{align}
    & \ket{\phi_{i^m}} = \ket{\psi_{i^m}} + \tau_{i^m}^{a^m}\ket{\psi_{a^m}}\,,\quad \hat{a}^\dagger_{i^m} = \hat{c}_{i^m}^\dagger + \tau_{i^m}^{a^m}\hat{c}^\dagger_{a^m}\,, \\
    & \bra{\Tilde{\phi}_{a^m}} = \bra{\psi_{a^m}} - \tau_{i^m}^{a^m}\bra{\psi_{i^m}}\,,\quad \Tilde{\hat{a}}_{a^m} = \hat{c}_{a^m} - \tau_{i^m}^{a^m}\hat{c}_{i^m}\,,\\
    & \ket{\phi_{a^m}} = \ket{\psi_{a^m}}\,,\quad \hat{a}^\dagger_{a^m} = \hat{c}_{a^m}^\dagger \,,\\
    &\bra{\Tilde{\phi}_{i^m}} = \bra{\psi_{i^m}}\,, \quad  \Tilde{\hat{a}}_{i^m} = \hat{c}_{i^m} \,,\\
    & \Tilde{\tau}_I^A = \tau_I^A\,,\quad \Tilde{\lambda}_A^I = \lambda_A^I \,,\quad \ket{\Tilde{\Psi}_{\rm R}} = \ket{\Psi_{\rm R}}\,,\quad \bra{\Tilde{\Psi}_{\rm L}} = \bra{\Psi_{\rm L}}\,.
\end{align}
These identities are ensured by the invariance of particle-hole orbital rotation in OATDCC parameterization: 
\begin{equation}
    \begin{aligned}
        &\hat{a}_{p^m}^\dagger \rightarrow e^{\Tilde{\hat{T}}_1}\hat{a}_{p^m}^\dagger  e^{-\Tilde{\hat{T}}_1}\,, \quad \Tilde{\hat{a}}_{p^m} \rightarrow e^{\Tilde{\hat{T}}_1}\Tilde{\hat{a}}_{p^m}  e^{-\Tilde{\hat{T}}_1}\,, \\
        & \bra{\Tilde{\Phi}} \rightarrow \bra{\Tilde{\Phi}} e^{-\Tilde{\hat{T}}_1} = \bra{\Tilde{\Phi}}\,,\quad \ket{{\Phi}} \rightarrow e^{\Tilde{\hat{T}}_1}\ket{{\Phi}} \,,\\
        & \bra{\Tilde{\Psi}_{\rm L}} \rightarrow \bra{\Tilde{\Phi}}(1+e^{\Tilde{\hat{T}}_1}\Tilde{\hat{\Lambda}}e^{-\Tilde{\hat{T}}_1})e^{-\Tilde{\hat{T}} + \Tilde{\hat{T}}_1} = \bra{\Tilde{\Psi}_{\rm L}} \,, \\
        &\ket{\Tilde{\Psi}_{\rm R}} \rightarrow e^{\Tilde{\hat{T}} - \Tilde{\hat{T}}_1}e^{\Tilde{\hat{T}}_1}\ket{\Tilde{\Phi}} = \ket{\Tilde{\Psi}_{\rm R}}\,.
    \end{aligned}
\end{equation}
where $\Tilde{\hat{T}}_1 = \Tilde{\tau}_{i^m}^{a^m}\Tilde{\hat{E}}_{i^m}^{a^m}$ is an arbitrary single excitation operator. Such a rotation invariance causes the $\Tilde{\hat{T}}_1$ redundancy in OATDCC and has been extensively discussed in the supplementary material of the original OATDCC paper\cite{Kvaal:2012}.

To convert sOATDCC wavefunctions to TD-OCCT1 correspondence needs two steps. The first step is to orthonormalize $\ket{\phi_{a^m}}$ and $\bra{\Tilde{\phi}_{i^m}}$,
\begin{equation}
        \ket{\phi_{a^m}^\prime} = G_{b^m}^{a^m}\ket{\phi_{b^m}}\,,\quad \bra{\Tilde{\phi}_{i^m}^\prime}= \Tilde{G}_{j^m}^{i^m}\bra{\Tilde{\phi}_{j^m}}\,,
\end{equation}
where $\ket{\phi_{a^m}^\prime}$ and $\bra{\Tilde{\phi}_{i^m}^\prime}$ are orthonormalized vectors, $G$ and $\Tilde{G}$ are transformation matrices. We also denote quantities after the orthonormalization as original quantities with primes. Dual vectors of $\ket{\phi_{a^m}^\prime}$ and $\bra{\Tilde{\phi}_{i^m}^\prime}$ are
\begin{equation}
        \ket{\phi_{i^m}^\prime} = (\Tilde{G}^{-1})_{i^m}^{j^m}\ket{\phi_{j^m}}\,,\quad \bra{\Tilde{\phi}_{a^m}^\prime}= ({G}^{-1})_{a^m}^{b^m}\bra{\Tilde{\phi}_{b^m}}\,.
\end{equation}
We also define general-rank transformations of $G$ and $\Tilde{G}$ as $G_I^J$ and $\Tilde{G}_A^B$, which are the tensor products of single-rank transformations of $G_{i^m}^{j^m}$ and $\Tilde{G}_{a^m}^{b^m}$, respectively. Wavefunctions as well as relevant quantities under such a transformation are given by
\begin{equation}
    \begin{aligned}
        &\Tilde{\hat{E}}_I^A\rightarrow(\Tilde{\hat{E}}_I^A)^\prime = G^A_B\Tilde{\hat{E}}_J^B\Tilde{G}^I_J\,,\\ &\Tilde{\hat{E}}_A^I\rightarrow(\Tilde{\hat{E}}_A^I)^\prime =  (\Tilde{G}^{-1})_I^J \Tilde{\hat{E}}_B^J({G}^{-1})_A^B \,,\\
&\Tilde{{\tau}}_I^A\rightarrow(\Tilde{\tau}_I^A)^\prime = (G^{-1})_A^B\Tilde{\tau}_J^B(\Tilde{G}^{-1})_I^J\,,\\ &\Tilde{{\lambda}}_A^I\rightarrow(\Tilde{\lambda}_A^I)^\prime =  \Tilde{G}^I_J \Tilde{\lambda}_B^J{G}_B^A \,,\\
&\ket{\Tilde{\Psi}_{\rm R}}\rightarrow\ket{\Tilde{\Psi}_{\rm R}^\prime} = \prod_m(\det{(\Tilde{G}_{j^m}^{i^m})})^{-1}\ket{\Tilde{\Psi}_{\rm R}}\,,\\
&\bra{\Tilde{\Psi}_{\rm L}} \rightarrow \bra{\Tilde{\Psi}_{\rm L}^\prime} = \prod_m\det{(\Tilde{G}_{j^m}^{i^m})}\bra{\Tilde{\Psi}_{\rm L}}\,.
    \end{aligned}
\end{equation}
In fact, if one replaces single-rank $G$ and $\Tilde{G}$ with arbitrary invertible matrices, the above transformation invariance still holds and that is the reason of the redundancy of $\Tilde{X}_{i^m}^{j^m}$ and $\Tilde{X}_{a^m}^{b^m}$ \cite{Kvaal:2012}. 

The second step is to express $\ket{{\Psi}_{\rm R}}$ and $\bra{{\Psi}_{\rm L}}$ via $\ket{\Tilde{\Psi}_{\rm R}^\prime}$ and $\bra{\Tilde{\Psi}_{\rm L}^\prime}$,
\begin{align}
    &   \ket{\psi_{a^m}} = \ket{\phi_{a^m}}\,,\quad  \hat{c}_{a^m}^\dagger =(\hat{a}^\dagger_{a^m})^\prime \,, \\
    &   \bra{\psi_{i^m}} = \bra{\Tilde{\phi}_{i^m}}  \,,\quad  \hat{c}_{i^m} = (\Tilde{\hat{a}}_{i^m})^\prime  \,,\\
    &  \tau_I^A = (\Tilde{\tau}_I^A )^\prime \,,\quad  \lambda_A^I =(\Tilde{\lambda}_A^I )^\prime\,,\quad \tau_{i^m}^{a^m} = \braket{\phi_{a^m}^\prime|\phi_{i^m}^\prime}\,,\\
    &\ket{\Psi_{\rm R}} = \ket{\Tilde{\Psi}_{\rm R}^\prime} \,,\quad  \bra{\Psi_{\rm L}} = \bra{\Tilde{\Psi}_{\rm L}^\prime} \,.
\end{align}

Now we have finished the proof of the equivalence of sOATDCC and TD-OCCT1. Although it is unnecessary for the proof of the equivalence of two methods, we will present the explicit form of the transformation and show that EoMs of two methods coincide for the scenario that converting TD-OCCT1 to sOATDCC since it will give readers a straightforward understanding. Useful transformations are
\begin{equation}
\begin{aligned}
        &\Tilde{\hat{E}}_{i^m}^{a^m} = \hat{E}_{i^m}^{a^m}\,,\quad \Tilde{X}_{i^m}^{a^m} = X_{i^m}^{a^m} + \dot{\tau}_{i^m}^{a^m} - \tau_{j^m}^{a^m}\tau_{i^m}^{b^m}X^{j^m}_{b^m}\,,\\
        &\Tilde{\hat{E}}_{i^m}^{j^m} = \hat{E}_{i^m}^{j^n} + \tau_{j^m}^{a^m}\hat{E}_{i^m}^{a^m}\,, \quad \Tilde{X}_{i^m}^{j^m} = X_{a^m}^{j^m}\tau_{i^m}^{a^m}\,,\\
        &\Tilde{\hat{E}}_{a^m}^{b^m} = \hat{E}_{a^m}^{b^n} - \tau_{i^m}^{a^m}\hat{E}_{i^m}^{b^m}\,,\quad \Tilde{X}_{a^m}^{b^m} = -\tau_{i^m}^{b^m}X^{i^m}_{a^m}\,,\\
        &\Tilde{\hat{E}}^{i^m}_{a^m} = \hat{E}^{i^m}_{a^m} - \tau_{j^m}^{a^m}\hat{E}_{j^m}^{i^m} + \tau_{i^m}^{b^m}\hat{E}_{a^m}^{b^m} - \tau_{j^m}^{a^m}\tau_{i^m}^{b^m}\hat{E}_{j^m}^{b^m}\,,\\
        &\Tilde{\hat{E}}_{\alpha^m}^{i^m} = \hat{E}_{\alpha^m}^{i^m} + \tau_{i^m}^{a^m}\hat{E}_{\alpha^m}^{a^m}\,,\quad \Tilde{X}_{\alpha^m}^{i^m} = X_{\alpha^m}^{i^m}\,,\quad \Tilde{X}^{i^m}_{a^m} = X^{i^m}_{a^m}\,,\\
        &\Tilde{\hat{E}}_{\alpha^m}^{a^m} = \hat{E}_{\alpha^m}^{a^m} \,,\quad \Tilde{X}_{\alpha^m}^{a^m} = X_{\alpha^m}^{a^m} - \tau_{i^m}^{a^m}{X}_{\alpha^m}^{i^m}\,,\\
        &\Tilde{\hat{E}}^{\alpha^m}_{i^m} = \hat{E}^{\alpha^m}_{i^m} \,,\quad \Tilde{X}^{\alpha^m}_{i^m} = X^{\alpha^m}_{i^m} + \tau_{i^m}^{a^m}{X}^{\alpha^m}_{a^m}\,,\\
        &\Tilde{\hat{E}}^{\alpha^m}_{a^m} = \hat{E}^{\alpha^m}_{a^m} - \tau_{i^m}^{a^m}\hat{E}^{\alpha^m}_{i^m}\,,\quad \Tilde{X}^{\alpha^m}_{a^m} = X^{\alpha^m}_{a^m} \,,\\
        &\Tilde{\hat{X}} = \hat{X} + (\dtp{\hat{T}_1})\,.
\end{aligned}
\end{equation}
where we already set $X_{i^m}^{j^m}$ and $X_{a^m}^{b^m}$ as zero. We stress that the constraint selection of $\Tilde{X}_{i^m}^{j^m}$ and $\Tilde{X}_{a^m}^{b^m}$ in sOATDCC in this case is no longer zero, which is not a usual choice in practical simulations. It is straightforward to verify that EoMs of active-virtual orbital rotations and CC coefficients are satisfied. To show the EoMs of intergroup rotations of active orbitals also hold, one notices that their EoMs can be expressed uniformly as 
\begin{equation}
    \begin{aligned}
        &\langle\Tilde{\Psi}_{\rm L}|[\hat{H}-i(\dtp\Tilde{\hat{T}})-i\Tilde{\hat{X}},\Tilde{\hat{E}}_{p^m}^{q^m}]|\Tilde{\Psi}_{\rm R}\rangle  \\
+&i\langle\Tilde{\Phi}|(\dtp\Tilde{\hat{\Lambda}})e^{-\Tilde{\hat{T}}}\Tilde{\hat{E}}_{p^m}^{q^m}e^{\Tilde{\hat{T}}}|\Tilde{\Phi}\rangle = 0\,,
    \end{aligned}
\end{equation}
which also holds since it is the summation (with some weights) of 
\begin{equation}
    \begin{aligned}
        &\langle\Psi_{\rm L}|[\hat{H}-i(\dtp\hat{T})-i\hat{X},\hat{E}^{q^m}_{p^m}]|\Psi_{\rm R}\rangle \\
+&i\langle\Phi|(\dtp\hat{\Lambda})e^{-\hat{T}}\hat{E}^{q^m}_{p^m}|\Psi_{\rm R}\rangle = 0\,.
\end{aligned}
\end{equation}

\section{EoMs of TD-OCCH}\label{Sec.TD-OCCH}
In this section, we will present the EoMs of single amplitudes and particle-hole orbital rotations of TD-OCCH. EoMs of other parameters of the method have already been presented in the Sec.~\ref{Sec.EoMs}. We will use notations $m_{\rm O}$, $m_{\rm T}$, $m_{\rm B}$, and $m_{\rm X}$ to label the fermion kinds with parameterization of TD-OCC, TD-OCCT1, TD-BCC, and TD-OCCX0. $m$ without any subscripts represent an arbitrary fermion kind. $X_{i^{m_{\rm X}}}^{a^{m_{\rm X}}}$ are set as Eq.~(\ref{eq.VGLG}) in VG and zero in LG. EoMs of $\lambda^{i^{m_{\rm X}}}_{a^{m_{\rm X}}}$ and $X_{i^{m_{\rm T}}}^{a^{m_{\rm T}}}$ are decoupled with other EoMs

\begin{equation}
i\left( \rho_{j^{m_{\rm T}}}^{i^{m_{\rm T}}}\delta_{b^{m_{\rm T}}}^{a^{m_{\rm T}}} - \rho_{a^{m_{\rm T}}}^{b^{m_{\rm T}}}\delta_{j^{m_{\rm T}}}^{i^{m_{\rm T}}} \right)X_{b^{m_{\rm T}}}^{j^{m_{\rm T}}} =
\bra{\Psi_{\rm L}} [\hat{H},\hat{E}_{i^{m_{\rm T}}}^{a^{m_{\rm T}}}] \ket{\Psi_{\rm R}}\,, 
\end{equation}

\begin{equation} 
\begin{aligned}
&-i\dot{\lambda}^{i^{m_{\rm X}}}_{a^{m_{\rm X}}}= \\
&\bra{\Psi_{\rm L}} [\hat{H},\hat{E}_{i^{m_{\rm X}}}^{a^{m_{\rm X}}}] \ket{\Psi_{\rm R}} - 
i\left( \rho_{j^{m_{\rm X}}}^{i^{m_{\rm X}}}\delta_{b^{m_{\rm X}}}^{a^{m_{\rm X}}} - \rho_{a^{m_{\rm X}}}^{b^{m_{\rm X}}}\delta_{j^{m_{\rm X}}}^{i^{m_{\rm X}}} \right)X_{b^{m_{\rm X}}}^{j^{m_{\rm X}}}\,.     
\end{aligned}
\end{equation}

After obtaining $\dot{\lambda}^{i^{m_{\rm X}}}_{a^{m_{\rm X}}}$ and $X_{i^{m_{\rm T}}}^{a^{m_{\rm T}}}$, one can solve the following coupled equations to get $\dot{\lambda}^{i^{m_{\rm B}}}_{a^{m_{\rm B}}}$, $X_{i^{m_{\rm B}}}^{a^{m_{\rm B}}}$, and $X_{i^{m_{\rm O}}}^{a^{m_{\rm O}}}$.
\begin{widetext}
    \begin{align}
&i\frac{1}{2}\left\{\sum_{n_{\rm B}}\langle\Phi_{j^{n_{\rm B}}}^{b^{n_{\rm B}}}| e^{-\hat{T}} \hat{E}^{i^{m_{\rm B}}}_{a^{m_{\rm B}}}|\Psi_{\rm R}\rangle\dot{\lambda}^{j^{n_{\rm B}}}_{b^{n_{\rm B}}} + (\dot{\lambda}^{i^{m_{\rm B}}}_{a^{m_{\rm B}}})^* + \sum_{n\in n_{\rm B},n_{\rm O}}A^{a^{m_{\rm B}}j^n}_{i^{m_{\rm B}}b^n}  (X^{b^n}_{j^n})^* \right\} + i(\delta_{b^{m_{\rm B}}}^{a^{m_{\rm B}}}D_{i^{m_{\rm B}}}^{j^{m_{\rm B}}}-D_{b^{m_{\rm B}}}^{a^{m_{\rm B}}}\delta_{i^{m_{\rm B}}}^{j^{m_{\rm B}}})X_{j^{m_{\rm B}}}^{b^{m_{\rm B}}}= \nonumber \\
&B_{i^{m_{\rm B}}}^{a^{m_{\rm B}}} -i \sum_{n\in n_{\rm T},n_{\rm X}}\frac{1}{2}A^{a^{m_{\rm B}}j^n}_{i^{m_{\rm B}}b^n}  (X^{b^n}_{j^n})^* 
-i\frac{1}{2}\sum_{n_{\rm X}}\langle\Phi_{j^{n_{\rm X}}}^{b^{n_{\rm X}}}| e^{-\hat{T}} \hat{E}^{i^{m_{\rm B}}}_{a^{m_{\rm B}}}|\Psi_{\rm R}\rangle\dot{\lambda}^{j^{n_{\rm X}}}_{b^{n_{\rm X}}}\,,  
\end{align}

    \begin{align}
&i\frac{1}{2}\left\{\sum_{n_{\rm B}}\langle\Phi_{j^{n_{\rm B}}}^{b^{n_{\rm B}}}| e^{-\hat{T}} \hat{E}^{i^{m_{\rm O}}}_{a^{m_{\rm O}}}|\Psi_{\rm R}\rangle\dot{\lambda}^{j^{n_{\rm B}}}_{b^{n_{\rm B}}} + \sum_{n\in n_{\rm B},n_{\rm O}}A^{a^{m_{\rm O}}j^n}_{i^{m_{\rm O}}b^n}  (X^{b^n}_{j^n})^* \right\} + i(\delta_{b^{m_{\rm O}}}^{a^{m_{\rm O}}}D_{i^{m_{\rm O}}}^{j^{m_{\rm O}}}-D_{b^{m_{\rm O}}}^{a^{m_{\rm O}}}\delta_{i^{m_{\rm O}}}^{j^{m_{\rm O}}})X_{j^{m_{\rm O}}}^{b^{m_{\rm O}}}= \nonumber \\
&B_{i^{m_{\rm O}}}^{a^{m_{\rm O}}} -i \sum_{n\in n_{\rm T},n_{\rm X}}\frac{1}{2}A^{a^{m_{\rm O}}j^n}_{i^{m_{\rm O}}b^n}  (X^{b^n}_{j^n})^* 
-i\frac{1}{2}\sum_{n_{\rm X}}\langle\Phi_{j^{n_{\rm X}}}^{b^{n_{\rm X}}}| e^{-\hat{T}} \hat{E}^{i^{m_{\rm O}}}_{a^{m_{\rm O}}}|\Psi_{\rm R}\rangle\dot{\lambda}^{j^{n_{\rm X}}}_{b^{n_{\rm X}}}\,,  
\end{align}

\begin{align}
 &iX_{i^{m_{\rm B}}}^{a^{m_{\rm B}}} + \sum_{n\in n_{\rm B},n_{\rm O}}i\langle\Phi_{i^{m_{\rm B}}}^{a^{m_{\rm B}}}| e^{-\hat{T}} \hat{E}^{j^n}_{b^n}|\Psi_{\rm R}\rangle X^{j^n}_{b^n} =  \langle\Phi_{i^{m_{\rm B}}}^{a^{m_{\rm B}}}| e^{-\hat{T}} \hat{H}|\Psi_{\rm R}\rangle -  \sum_{n\in n_{\rm T},n_{\rm X}}i\langle\Phi_{i^{m_{\rm B}}}^{a^{m_{\rm B}}}| e^{-\hat{T}} \hat{E}^{j^n}_{b^n}|\Psi_{\rm R}\rangle X^{j^n}_{b^n}\,.
\end{align}

\end{widetext}

Finally, $\dot{\tau}_{i^{m_{\rm T}}}^{a^{m_{\rm T}}}$ and $\dot{\tau}_{i^{m_{\rm X}}}^{a^{m_{\rm X}}}$ can be solved

  \begin{align}
& i(\delta_{b^{m_{\rm T}}}^{a^{m_{\rm T}}}D_{i^{m_{\rm T}}}^{j^{m_{\rm T}}}-D_{b^{m_{\rm T}}}^{a^{m_{\rm T}}}\delta_{i^{m_{\rm T}}}^{j^{m_{\rm T}}})\dot{\tau}_{j^{m_{\rm T}}}^{b^{m_{\rm T}}}= B_{i^{m_{\rm T}}}^{a^{m_{\rm T}}}\nonumber \\
 -&i(\delta_{b^{m_{\rm T}}}^{a^{m_{\rm T}}}D_{i^{m_{\rm T}}}^{j^{m_{\rm T}}}-D_{b^{m_{\rm T}}}^{a^{m_{\rm T}}}\delta_{i^{m_{\rm T}}}^{j^{m_{\rm T}}})X_{j^{m_{\rm T}}}^{b^{m_{\rm T}}} - \sum_n\frac{i}{2}A^{a^{m_{\rm T}}j^n}_{i^{m_{\rm T}}b^n}  (X^{b^n}_{j^n})^*\nonumber\\
 -&\frac{i}{2}\sum_{n\in n_{\rm B},n_{\rm X}}\langle\Phi_{j^{n}}^{b^{n}}| e^{-\hat{T}} \hat{E}^{i^{m_{\rm T}}}_{a^{m_{\rm T}}}|\Psi_{\rm R}\rangle\dot{\lambda}^{j^{n}}_{b^{n}}\,,  
\end{align}

\begin{equation}
    i\dot{\tau}_{i^{m_{\rm X}}}^{a^{m_{\rm X}}} =  \bra{\Phi_{i^{m_{\rm X}}}^{a^{m_{\rm X}}}}e^{-\hat{T}}(\hat{H} - i\hat{X})\ket{\Psi_{\rm R}}\,.
\end{equation}
It is straightforward to verify that TD-OCCH reduce to other four methods when the corresponding parameterization is used in all fermion kinds. It is also possible to use two or three types of parameterization in TD-OCCH method, and one needs to remove unused types of parameters in above EoMs. An attractive parameterization is the OCCT1-OCCX0 mixed one, in which all single amplitudes and particle-hole rotations of different kinds are decoupled. One can also assign few kinds fermions OCC and BCC type parameterization, in which only few of coupled equations need to be solved.

Similar to the TD-OCCX0 method, the non-variance of $X_{i^{m_{\rm X}}}^{a^{m_{\rm X}}}$ makes the Ehrenfest theorem\cite{Sato:2016} modified even without frozen-cores. One only needs to account for the TD-OCCX0-type parameterization fermion kinds
\begin{equation}\label{eq:mEhrenMix}
\begin{aligned}
         2i\frac{d}{dt}\braket{\hat{O}} &= \braket{\Psi_{\rm L}|[\hat{O},\hat{H}]|\Psi_{\rm R}} + \braket{\Psi_{\rm R}|[\hat{O},\hat{H}]|\Psi_{\rm L}} \\
        &+ \sum_{m\in m_{\rm X}} O_{t^m}^{u^m}\{\braket{\Psi_{\rm L}|(\hat{H}-i\hat{X})e^{\hat{T}}\bar{\hat{\Pi}}e^{-\hat{T}}\hat{E}_{t^m}^{u^m}|\Psi_{\rm R}}  \\
        &-\braket{\Psi_{\rm L}|\hat{E}_{t^m}^{u^m}e^{\hat{T}}\bar{\hat{\Pi}}e^{-\hat{T}}(\hat{H}-i\hat{X})|\Psi_{\rm R}} \\
        &+ \braket{\Psi_{\rm R}|(\hat{H}-i\hat{X})e^{\hat{T}}\bar{\hat{\Pi}}e^{-\hat{T}}\hat{E}_{t^m}^{u^m}|\Psi_{\rm L}}  \\
        &-\braket{\Psi_{\rm R}|\hat{E}_{t^m}^{u^m}e^{\hat{T}}\bar{\hat{\Pi}}e^{-\hat{T}}(\hat{H}-i\hat{X})|\Psi_{\rm L}} \}\,.
\end{aligned}
\end{equation}
When frozen orbitals are included, two additional terms that replace $t^m$, $u^m$ with $i^{\prime\prime m}$, $\mu^m$ and $\mu^m$, $i^{\prime\prime m}$ in the last term of Eq.~(\ref{eq:mEhrenMix}) should be added. Additionally, the summation for the new terms should be for all kinds of fermions.

\bibliography{refs.bib}

\begin{thebibliography}{78}
\bibitem [{TDH()}]{TDH-TDMVCC}%
  \BibitemOpen
  \href@noop {} {}\bibinfo {note} {The truncation in TDMVCC is a TDH-TDMVCC hybrid form, which is equivalent to setting the weights of the TDH kinds as infinity}\BibitemShut {NoStop}%
\end{thebibliography}

\end{document}